\documentclass[paper]{JHEP3}

\usepackage{amsmath,amssymb}

\newcommand{\be}{\begin{equation}}
\newcommand{\ee}{\end{equation}}
\newcommand{\lb}[1]{\label{#1}}

\renewcommand{\theequation}{\arabic{section}.\arabic{equation}}
\newcommand{\p}[1]{(\ref{#1})}
\def\tr{{\rm tr}\,}
\def\cN{{\cal N}}
\def\cD{{\cal D}}
\def\tU{\textrm{U}}
\def\bea{\begin{eqnarray}}
\def\eea{\end{eqnarray}}
\def\nn{\nonumber}

\def\un{\underline}
\def\theequation{\arabic{section}.\arabic{equation}}

\preprint{ITP--UH--22/08\\
JINR-E2-2008-175}

\title{ABJM models in $\cN=3$ harmonic superspace}

\author{I.L. Buchbinder\\
    Department of Theoretical Physics, Tomsk State
Pedagogical University, 634041 Tomsk, Russia,
    E-mail: \email{joseph@tspu.edu.ru}}

\author{E.A. Ivanov\\
Bogoliubov Laboratory of Theoretical Physics, JINR, 141980 Dubna,
Russia \\
E-mail: \email{eivanov@theor.jinr.ru}}

\author{O. Lechtenfeld\\
Institut f\" ur Theoretische Physik,
Leibniz Universit\" at Hannover, 30167 Hannover, Germany,
E-mail: \email{lechtenf@itp.uni-hannover.de}}

\author{N.G. Pletnev\\
Department of Theoretical Physics, Institute of Mathematics,
Novosibirsk, 630090, Russia,
E-mail: \email{pletnev@math.nsc.ru}}

\author{I.B. Samsonov\\
Laboratory of Mathematical Physics, Tomsk Polytechnic
University, 634050 Tomsk, Russia
$\&$ Istituto Nazionale di Fisica Nucleare, Sezione di Padova, 35131 Padova,
Italy\\
E-mail: \email{samsonov@mph.phtd.tpu.ru}}

\author{B.M. Zupnik\\
Bogoliubov Laboratory of Theoretical Physics, JINR, 141980 Dubna, Russia \\
E-mail: \email{ zupnik@theor.jinr.ru}}

\abstract{We construct the classical action of the
Aharony-Bergman-Jafferis-Maldacena (ABJM) model in the $\cN{=}3$, $d{=}3$
harmonic superspace. In such a formulation three out of six supersymmetries
are realized off shell while the other three mix the
superfields and close on shell. The superfield
action involves two hypermultiplet superfields in the
bifundamental representation of the gauge group and
two Chern-Simons gauge superfields corresponding to the left and
right gauge groups. The $\cN{=}3$ superconformal invariance allows only
for a minimal gauge interaction of the hypermultiplets.
Amazingly, the correct sextic scalar potential of ABJM emerges
after the elimination of auxiliary fields.
Besides the original $\tU(N)\times \tU(N)$ ABJM model, we also construct
$\cN{=}3$ superfield formulations of some generalizations. For
the $\textrm{SU}(2)\times \textrm{SU}(2)$ case we give a simple superfield
proof of its enhanced $\cN{=}8$ supersymmetry and $\textrm{SO}(8)$ R-symmetry.}


\begin{document}
\setcounter{equation}{0}
\section{Introduction}

The last year has witnessed impressive progress in constructing
the actions of multiple M2 branes and studying their properties.
M2 branes can be described by three-dimensional superconformal field theories,
which have the structure of Chern-Simons-matter theory
with $\cN{=}6$ or $\cN{=}8$ extended supersymmetry.
The problem of constructing such actions for multiple M2
branes was raised several years ago in~\cite{Schwarz04},
but was resolved only recently in a
series of works \cite{BLG,BLS,Gustavsson,ABJM,BLS1}. Various
aspects of these theories were studied subsequently;
a partial list of papers is \cite{Benna}--\cite{M2D2}.

Of special interest is the work of Aharony, Bergman,
Jafferis and Maldacena (ABJM) \cite{ABJM} in which the
three-dimensional $\cN{=}6$ superconformal theory was constructed
and proved to describe multiple M2 branes on the ${\mathbb
C}^4/{\mathbb Z}_k$ orbifold. The ABJM model plays a fundamental
role, since many three-dimensional superconformal theories such as
the Bagger-Lambert-Gustavsson (BLG) model with maximal $\cN{=}8$
supersymmetry~\cite{BLG,Gustavsson} and other models with less
supersymmetry follow from the ABJM one under particular choices
of the gauge group.
The field content of the ABJM model is given by four complex
scalar and spinor fields which live in the bifundamental
representation of the $\tU(N)\times \tU(N)$ gauge group \footnote{
Generalizations to some other gauge groups are described, e.g.,
in~\cite{Hosomichi,STac,ABJ,Li}.} while the gauge fields are governed by
Chern-Simons actions of levels $k$ and $-k$, respectively.

It is desirable to have a superfield description of the ABJM models,
with maximal number of manifest and off-shell supersymmetries.
As in other cases, such superfield formulations are expected to bring to light
geometric and quantum properties of the theory which are implicit
in the component formulation.
To date, several approaches to the superfield description of the ABJM
and BLG theories are known.  They use either $\cN{=}1$ and $\cN{=}2$
off-shell superfields~\cite{Benna,Petkou,Cherkis} or $\cN{=}6$ and $\cN{=}8$
on-shell superfields~\cite{Cederwall,Bandos}. These
formulations were able to partly clarify the origin of the interaction
of scalar and spinor component fields \footnote{
Earlier important references on superfield extensions of Chern-Simons theory
with and without matter couplings are \cite{Si}, \cite{ZP1} and \cite{ZK}.}.

In the present paper we take the next step in working out off-shell
superfield formulations of the ABJM theory. Namely, we develop its formulation
in $\cN{=}3, d{=}3$ harmonic superspace, which was proposed in~\cite{ZK,Z3}
as the appropriate adaptation of the $\cN{=}2, d{=}4$ harmonic superspace
\cite{GIKOS,book}. The four complex scalars and spinors are embedded
into two $q$~hypermultiplet analytic superfields which sit in the bifundamental
representation of the $\tU(N)\times \tU(N)$ gauge group. The gauge
part of the action is given by a sum of two $\cN{=}3$ supersymmetric
Chern-Simons actions with levels $k$ and~$-k$, respectively,
just as in the component approach~\cite{ABJM}. In this formulation,
three out of six supersymmetries are realized off shell and are manifest,
while the other three transform the gauge superfields and hypermultiplets
into each other and close only on shell. The same concerns the full
automorphism group $\textrm{SO}(6)\sim\textrm{SU}(4)$ of the $\cN{=}6$
supersymmetry: only its $\textrm{SU}(2)\times\textrm{SU}(2)$ subgroup
is manifest in the $\cN{=}3$ superfield formalism,
while the coset $\textrm{SU}(4)/[\textrm{SU}(2)\times\textrm{SU}(2)]$
is realized by nonlinear superfield transformations with an on-shell
closure \footnote{
The pure Chern-Simons theory also admits off-shell $\cN{=}5$ and $\cN{=}6$
extensions in some specific harmonic superspaces~\cite{HL,Z8}. However, it is
likely that analogous superextensions of the Chern-Simons-matter systems do not
exist, rendering the $\cN{=}3$ extension as the {\it maximal off-shell\/} one.}.

The scale invariance of the ABJM theory imposes severe
restrictions on the action in the $\cN{=}3$ superfield formulation:
only minimal interactions of the $q$~hypermultiplets with the
gauge superfields are admissible, and no explicit superpotential can be
constructed. One may wonder how the sextic scalar potential of
the ABJM model can appear in the absence of an original superpotential.
We show that, upon reducing the superfield action
to the component form, the scalar potential naturally arises
as a result of eliminating some auxiliary fields from the gauge multiplet
and from the harmonic expansion of the off-shell $q$~hypermultiplets.
This is a striking new feature of the $\cN{=}3$ superfield formulation as
compared to the $\cN{=}1$ and $\cN{=}2$ ones.

The paper is organized as follows. In Section~2 we
collect the basic building blocks of the $\cN{=}3$, $d{=}3$ harmonic
superspace approach which are used in Section~3 for
constructing the $\cN{=}3$ superfield action of the ABJM model and for
demonstrating its $\cN{=}6$ and $\textrm{SO}(6)$ (super)invariances,
for the gauge group $\tU(N)\times \tU(M)$. We also show how the sextic scalar
potential of the ABJM model emerges.
In Section~4 we present $\cN{=}3$ superfield formulations for a variant of
the ABJM theory with gauge group $\textrm{SO}(N)\times \textrm{USp}(2M)$,
which respects $\cN{=}5$ supersymmetry and $\textrm{SO}(5)$ R-symmetry. We also
demonstrate in a simple way that the $\textrm{SU}(N)\times \textrm{SU}(M)$
model admits hidden supersymmetry and R-symmetry
($\cN{=}6$ and $\textrm{SO}(6)$) only for the choice $N{=}M\,$.
Section~5 is devoted to the special $\textrm{SU}(2)\times \textrm{SU}(2)$ case
in which the ABJM model coincides with the BLG one.
We present in $\cN{=}3$ superfield form the hidden $\cN{=}8$ supersymmetry
and $\textrm{SO}(8)$ R-symmetry of this model. The final Section~6 contains
a discussion of our results and marks prospects of their applications to
M2 branes and their relation with D2 branes. In an Appendix, for the simple
example of the $\tU(1)\times \tU(1)$ model,
we describe the $\cN{=}3$ superfield realization of the Higgs-type effect
of~\cite{M2D2} which relates M2 branes to D2 branes.

\setcounter{equation}{0}
\section{Gauge and matter theories in $\cN{=}3\,,$ $d{=}3$
harmonic
superspace}
\subsection{Superspace conventions}
We start with a short review of the $\cN{=}3$, $d{=}3$ harmonic superspace
and field models therein which were originally introduced in
\cite{ZK,Z3}. Our three-dimensional notations are as follows:
we use the Greek letters $\alpha,\beta,\ldots$ to label the
spinorial indices corresponding to the $\textrm{SO}(1,2)\simeq \textrm{SL}(2,R)$ Lorentz group.
 A vector in $d{=}3$ Minkowski space
is equivalent to a second-rank symmetric spinor,
$x^{\alpha\beta}=x^m(\gamma_m)^{\alpha\beta}$
\bea
&&(\gamma_m)_\alpha^\rho(\gamma_n)_\rho^\beta=-(\gamma_m)_{\alpha\rho}
(\gamma_n)^{\rho\beta}
=-\eta_{mn}\delta^\beta_\alpha+\varepsilon_{mnp}(\gamma^p)^\beta_\alpha,\nn\\
&&(\gamma^m)^{\alpha\beta}(\gamma_m)_{\rho\sigma}= 2\delta^\alpha_{(\rho}
\delta^\beta_{\sigma)}\,,
\eea
where $\eta_{mn}=\mbox{diag}(1,-1,-1)$ is the $d{=}3$ Minkowski metric.
 The R-symmetry of $\cN{=}3$
superspace is $\textrm{SO}(3)_R\simeq \textrm{SU}(2)_R$. Therefore we label the three
copies of Grassmann variables by a pair of symmetric $\textrm{SU}(2)$
indices $i,j$, i.e.,
$\theta_\alpha^{ij}=\theta_\alpha^{ji}$. Hence, the $\cN{=}3$
superspace is parametrized by
the following real coordinates
\be z=(x^m,
\theta_\alpha^{ij}),\quad \overline{x^m}=x^m,\quad
 \overline{\theta_\alpha^{ij}}=\theta_{ij\alpha}\,.
\ee

The partial spinor and vector derivatives  are
defined as follows
\be
\frac\partial{\partial\theta^{ij}_\alpha}\theta^{kl}_\beta
=\delta^\alpha_\beta\,\delta^{k}_{(i}\delta^l_{j)}\,,\quad
\partial_{\alpha\beta} x^{\rho\sigma}
= 2 \delta^\rho_{(\alpha}\delta^\sigma_{\beta)}\,, \quad
\partial_{\alpha\beta}
=(\gamma^m)_{\alpha\beta}\frac{\partial}{\partial x^m}\,.
\ee
These
derivatives are used to construct covariant spinor
derivatives and supercharges,
\be
D^{kj}_\alpha=\frac\partial{\partial\theta^\alpha_{kj}}
 +i\theta^{kj\,\beta}\partial_{\alpha\beta},\quad
 Q^{kj}_\alpha=
 \frac\partial{\partial\theta^\alpha_{kj}}
 -i\theta^{kj\,\beta}\partial_{\alpha\beta}\,.
\label{Q}
\ee
The spinor indices as well as the R-symmetry ones are raised and
lowered with the antisymmetric two-dimensional tensors
$\varepsilon_{\alpha\beta}$, $\varepsilon_{ij}$, respectively ($\varepsilon_{12}=-\varepsilon^{12}=1\,$).

 We use standard harmonic variables $u^\pm_i$
parametrizing the coset $\textrm{SU}(2)/\tU(1)$ \cite{GIKOS,book}.
In particular, the partial harmonic derivatives  are
\be
\partial^{++}=u^+_i\frac\partial{\partial u^-_i},\quad
\partial^{--}=u^-_i\frac\partial{\partial u^+_i},\quad
\partial^0=[\partial^{++},\partial^{--}]=u^+_i\frac\partial{\partial u^+_i}
-u^-_i\frac\partial{\partial u^-_i}.
\ee
The harmonic projections of the Grassmann $\cN{=}3$ coordinates and spinor
derivatives can be defined as follows
\bea
\theta^{ij}_\alpha&\longrightarrow&
(\theta^{++}_\alpha,\theta^{--}_\alpha,\theta^0_\alpha)=
(u^+_iu^+_j\theta^{ij}_\alpha,u^-_iu^-_j\theta^{ij}_\alpha,
u^+_iu^-_j\theta^{ij}_\alpha),\nn\\
D^{ij}_\alpha&\longrightarrow&
(D^{++}_\alpha,D^{--}_\alpha,D^0_\alpha)=
(u^+_iu^+_jD^{ij}_\alpha,u^-_iu^-_jD^{ij}_\alpha,
u^+_iu^-_jD^{ij}_\alpha).
\eea

The analytic subspace in the full $\cN{=}3$ superspace
is parametrized by the
following coordinates:
\be
\zeta_A=(x^{\alpha\beta}_A,
\theta^{++}_\alpha, \theta^{0}_\alpha, u^\pm_i), \ee where \be
x^{\alpha\beta}_A=(\gamma_m)^{\alpha\beta}x^m_A=x^{\alpha\beta}
+i(\theta^{++\alpha}\theta^{--\beta}+\theta^{++\beta}\theta^{--\alpha}).
\ee
It is instructive to rewrite the harmonic and Grassmann
derivatives in the analytic coordinates,
\bea
{\cal
D}^{++}&=&\partial^{++}+2i\theta^{++ \alpha}\theta^{0 \beta}
 \partial^A_{\alpha\beta}
 +\theta^{++\alpha}\frac\partial{\partial\theta^{0 \alpha}}
 +2\theta^{0 \alpha}\frac\partial{\partial\theta^{--\alpha}},\nn\\
{\cal D}^{--}&=&\partial^{--}
 -2i\theta^{--\alpha}\theta^{0 \beta}\partial^A_{\alpha\beta}
 +\theta^{--\alpha}\frac\partial{\partial\theta^{0 \alpha}}
 +2\theta^{0 \alpha}\frac\partial{\partial\theta^{++ \alpha}},
\nn\\
{\cal D}^0&=&\partial^0+2\theta^{++ \alpha}\frac\partial{\partial\theta^{++ \alpha}}
-2\theta^{--\alpha}\frac\partial{\partial\theta^{--\alpha}}, \quad
[{\cal D}^{++}, {\cal D}^{--}]={\cal D}^0,
\eea
\be
D^{++}_\alpha=\frac{\partial}{\partial
\theta^{--\alpha}},\quad
D^{--}_\alpha=\frac\partial{\partial\theta^{++ \alpha}}
 +2i\theta^{--\beta}\partial^A_{\alpha\beta}, \quad
D^0_\alpha=-\frac12\frac\partial{\partial\theta^{0 \alpha}}
+i\theta^{0 \beta}\partial^A_{\alpha\beta}, \ee
where
$\partial^A_{\alpha\beta}=(\gamma^m)_{\alpha\beta}\partial/\partial
x^m_A
$. These derivatives satisfy the following
 relations:
 \be \{D^{++}_\alpha, D^{--}_\beta\}=2i\partial^A_{\alpha\beta}, \quad \{D^{0}_\alpha,
D^{0}_\beta\}=-i\partial^A_{\alpha\beta}, \quad\{D^{\pm\pm}_\alpha, D^{0}_\beta\} = 0\,,
\ee
\be
[{\cal
D}^{\mp\mp}, D^{\pm\pm}_\alpha]=2D^0_\alpha, \quad [{\cal D}^{0},
D^{\pm\pm}_\alpha]=\pm 2D^{\pm\pm}_\alpha, \quad [{\cal
D}^{\pm\pm}, D^0_\alpha]=D^{\pm\pm}_\alpha.
\label{D-alg}
\ee
The analytic
superfields are defined to be independent of the
$\theta^{--}_\alpha$ variable
\be
D^{++}_\alpha\Phi_A=0 \quad \Rightarrow \quad \Phi_A = \Phi_A(\zeta_A)\,.
\ee

We use the following conventions for the full and
analytic integration measures,
\bea
d^9z &=&-\frac1{16}d^3x
(D^{++})^2 (D^{--})^2(D^{0})^2,\\
d\zeta^{(-4)}&=&\frac14 d^3x_Adu (D^{--})^2(D^{0})^2\,, \quad d^9z du = -\frac{1}{4} d\zeta^{(-4)}(D^{++})^2\,, \lb{MeaS}
\eea
where $(D^{++})^2=D^{++\alpha}D^{++}_\alpha$ and similarly for
other objects. With such conventions, the superspace integration rules are most
simple:
\bea
\int
d\zeta^{(-4)}(\theta^{++})^2(\theta^0)^2 f(x_A)&=&\int d^3x_A
f(x_A)\,,\nn\\
\int d^9z(\theta^{++})^2(\theta^{--})^2(\theta^0)^2 f(x)&=&\int d^3x \,
f(x)\,,
\eea
for some field $f(x)$.

We denote the special conjugation in the $\cN{=}3$ harmonic superspace by
$\widetilde{\phantom{a}}$
\bea
\widetilde{(u^\pm_i)}=u^{\pm i},\quad \widetilde{(x^m_A)}=x^m_A,\quad \widetilde{(\theta^{\pm\pm}_\alpha)}=
\theta^{\pm\pm}_\alpha,\quad \widetilde{(\theta^0_\alpha)}=
\theta^0_\alpha.
\eea
It is squared to $-1$ on the harmonics  and to $1$ on $x^m_A$ and
 Grassmann coordinates.
All bilinear combinations of the Grassmann coordinates are imaginary
\bea
&&\widetilde{[(\theta^{++}_\alpha\theta^0_\beta)]} =-\theta^{++}_\alpha\theta^0_\beta,\quad
\widetilde{[(\theta^{++})^2]}=-(\theta^{++})^2,\quad
\widetilde{[(\theta^0)^2]}=-(\theta^0)^2.
\eea

The conjugation rules for the spinor and harmonic derivatives are
\bea
\widetilde{(D^0_\alpha\Phi)}=-D^0_\alpha\tilde\Phi,\quad
\widetilde{[(D^0)^2\Phi]}=-(D^0)^2\tilde\Phi,\quad \widetilde{(\cD^{++}\Phi)}=
\cD^{++}\tilde\Phi
\eea
where $\Phi$ and $\tilde\Phi$ are conjugated even superfields.
When the superfields are matrix-like objects, $\Phi=[\Phi^A_B]$,
the Hermitian conjugation assumes the $\widetilde{\phantom{m}}$ conjugation
and transposition, e.g.,
$[\Phi^A_B]^\dagger=\widetilde\Phi^B_A$.

The analytic superspace measure is real,
$\widetilde{d\zeta^{(-4)}}=d\zeta^{(-4)}$, while the full
superspace measure is imaginary,
$\widetilde{d^9z}=-d^9z$.

\subsection{Chern-Simons and hypermultiplet actions in
$\cN{=}3$ harmonic superspace}
\subsubsection{Chern-Simons action}
The $\cN{=}3$ supersymmetric gauge multiplet in three dimensions
consists of a triplet of real scalar fields $\phi^{(kl)}$, one
real vector $A_m$, real $\textrm{SU}(2)$-singlet spinor
$\lambda_\alpha$, $\textrm{SU}(2)$-triplet spinor $\chi_\alpha^{(kl)}$ and a
triplet of auxiliary fields $X^{(kl)}$. Altogether they constitute eight
bosonic and eight fermionic off-shell degrees of freedom. All
these components are embedded into an
analytic gauge superfield
which originally contains an infinite set of fields in its $\theta$ and $u$-expansion.
However, like in the $\cN{=}2\,,d{=}4$ case \cite{GIKOS,book},  the gauge freedom with an analytic  superfield parameter
allows one to
pass to the Wess-Zumino gauge which reveals the above finite irreducible
field content of the
$\cN{=}3$ gauge multiplet
\bea
V^{++}_{WZ}&=&3(\theta^{++})^2u^-_k u^-_l \phi^{kl}(x_A)
+2\theta^{++\alpha}\theta^{0\beta}A_{\alpha\beta}(x_A)
+2(\theta^0)^2\theta^{++\alpha}\lambda_\alpha(x_A)\nn\\&&
+3(\theta^{++})^2\theta^{0\alpha}u^-_k
u^-_l\chi_\alpha^{kl}(x_A) +3i(\theta^{++})^2(\theta^0)^2u^-_ku^-_l
X^{kl}(x_A).
\label{V++}
\eea
In the Abelian case the corresponding gauge transformation of the imaginary
superfield $V^{++}$ reads
\be
\delta_\Lambda V^{++}=-{\cal D}^{++}\Lambda,\qquad
\widetilde{\Lambda}=-\Lambda,
\ee
The non-Abelian gauge superfield has the following infinitesimal transformation law
\be \delta_\Lambda
V^{++}=-{\cal D}^{++}\Lambda-[V^{++},\Lambda].\lb{nonabV}
\ee
In what follows we shall be mainly interested in the gauge group $\tU(N)$ in the fundamental representation
and its adjoint. In this case, $V^{++}$ and $\Lambda$ are antihermitian $N\times N$ matrices
\be
\widetilde{[V^{++}{}^B_A]}=-V^{++}{}^A_B, \;\; \widetilde{[\Lambda^{B}_A]}=-\Lambda^{A}_B
\quad A, B=1, 2,\ldots ,N\,.
\ee
The $\textrm{SU}(N)$ case is singled out by the extra tracelessness condition
\be
V^{++}{}^A_A= \Lambda^A_A = 0\,.
\ee

Using $V^{++}$, one can construct either the
Yang-Mills or Chern-Simons actions in the $\cN{=}3$ superspace \cite{ZK,Z3}.
The non-Abelian Chern-Simons superfield action is
\be
 S_{CS}=\frac{ik}{4\pi}\,\tr\sum\limits^{\infty}_{n=2} \frac{(-1)^{n}}{n} \int
d^3x d^6\theta du_{1}\ldots du_n \frac{V^{++}(z,u_{1} ) \ldots
V^{++}(z,u_n )}{ (u^+_{1} u^+_{2})\ldots (u^+_n u^+_{1} )},
\label{CS}
\ee
where $k$ is the Chern-Simons level.
Note that this action is formally analogous to the superfield
action of the $\cN{=}2, d{=}4$ Yang-Mills theory \cite{Z2}, although
the full integration measure is $d^4x d^8\theta$ in the latter case.
The action (\ref{CS}) can be checked to be
invariant under the gauge transformation \p{nonabV}.

For what follows it will be necessary to know a general variation of the Chern-Simons
action (\ref{CS})
\be
\delta S_{CS}=-\frac{ik}{4\pi}\tr\int d^9z du\,\delta
V^{++} V^{--}\,.
\label{delta-CS1}
\ee
Here $V^{--}$ is the non-analytic gauge superfield which is related to $V^{++}$ by
the harmonic zero-curvature equation \cite{book,Z2}
\be
{\cal D}^{++}V^{--}-{\cal
D}^{--}V^{++}+[V^{++},V^{--}]=0
\label{zero-curv}
\ee
and is transformed under the gauge group as
\be \delta_\Lambda
V^{--}=-{\cal D}^{--}\Lambda-[V^{--},\Lambda]\,.\lb{nonabV-}
\ee
The
solution of (\ref{zero-curv}) is represented by the following
series
\be
V^{--}(z,u)=\sum_{n=1}^\infty (-1)^n \int du_1\ldots
du_n \frac{V^{++}(z,u_1)V^{++}(z,u_2)\ldots
V^{++}(z,u_n)}{(u^+u^+_1)(u^+_1u^+_2)\ldots (u^+_n u^+)}.
\label{V--}
\ee
The superfield $V^{--}$ can be used to define
the superfield strength $W^{++}$ \cite{Z3},
\be
W^{++}=-\frac14D^{++\alpha}D^{++}_\alpha V^{--},\qquad
{\cal D}^{++}W^{++}+[V^{++},W^{++}]=0.
\label{W}
\ee
By construction, $W^{++}$ is analytic and gauge
covariant, $\delta_\Lambda W^{++}= [\Lambda, W^{++}]$. Note
that $W^{++}$ is hermitian in contrast to the gauge superfield
$V^{++}$, $(W^{++})^\dagger= W^{++}$. In terms of $W^{++}$,
the variation (\ref{delta-CS1}) of the Chern-Simons
action can be written as
\be
\delta S_{CS}=-\frac{ik}{4\pi} \tr \int d\zeta^{(-4)} \delta V^{++}
W^{++}\,.
\label{var-CS}
\ee
The classical
equation of motion in the pure super Chern-Simons model is
$W^{++}=0$, which implies the superfields $V^{\pm\pm}$ to be pure gauge.
The topological character of the $\cN{=}3$ gauge multiplet with the Chern-Simons action (\ref{CS})
can also be seen directly from the component structure of this action \footnote{The component structure of the $\cN{=}3$ Chern-Simons action
with the matter couplings added was given in \cite{KL}.}:
\bea
S_{CS}&=&\frac{k}{4\pi}\tr\int d^3x\Big(
\phi^{kl} X_{kl}
-\frac{2i}3\phi^i_j[\phi^k_i,\phi^j_k]
+\frac i2\lambda^\alpha\lambda_\alpha
-\frac i4\chi^\alpha_{kl}\chi^{kl}_\alpha\nn\\&&
-\frac12A^{\alpha\beta}\partial^\gamma_{\alpha}A_{\beta\gamma}
-\frac{i}6
A^\alpha_\beta[A^\gamma_\alpha,A^\beta_\gamma] \Big).
\label{CS-comp}
\eea
Using $d{=}3$ $\gamma$-matrices
one can convert the vector part of the action (\ref{CS-comp}) to
the standard form $\varepsilon^{mnp}(A_m\partial_n
A_p-\frac{2i}3A_m A_nA_p)\,$.

As for the $\cN{=}3$, $d{=}3$ super Yang-Mills action, it is concisely written as the following integral
over the analytic superspace
\be
S_{SYM} = -\frac{1}{g^2}\, \tr \int d\zeta^{(-4)}\, (W^{++})^2\,, \;\; [g] = 1/2\,.\lb{3SYM}
\ee
It should be compared with the $\cN{=}2, d{=}4$ SYM action in the harmonic superspace which is represented either
by the action of the type \p{CS} or as an integral of the square of the relevant chiral (ant-chiral) superfield strength over
the chiral (anti-chiral) $\cN{=}2, d{=}4$ superspace \cite{GIKOS,book}.

\subsubsection{Hypermultiplet action}
Like in the $\cN{=}2$, $d{=}4$ case \cite{GIKOS,book}, the $\cN{=}3, d{=}3$ hypermultiplet is described
by an analytic harmonic superfield $q^+(\zeta)$ with the following free action
\be
S_q= \int d\zeta^{(-4)} \bar q^+{\cal D}^{++}q^+\,, \quad \bar q^+ = \widetilde{q^+}\,, \;\;
\widetilde{\bar q^+} = -q^+\,.\lb{qFree}
\ee
The physical fields of $\cN{=}3$, $d{=}3$ hypermultiplet are $\textrm{SU}(2)$ doublets
$f^i$ and $\psi^i_\alpha\,$.
After elimination of an infinite tower of
auxiliary fields by their equations of motion (they all vanish on shell)
the physical fields appear in the component expansion of the analytic
superfield $q^+$ and its
conjugated $\bar q^+$ as
\bea
q^+&=&u^+_i
f^i+(\theta^{++\alpha}
u^-_i-\theta^{0\alpha}u^+_i)\psi^i_\alpha
-2i(\theta^{++\alpha}\theta^{0\beta})\partial^A_{\alpha\beta}f^iu^-_i\,,
\nn\\
\bar q^+&=&-u^+_i \bar
f^i+(\theta^{++\alpha}
u^-_i-\theta^{0\alpha}u^+_i)\bar\psi^i_\alpha
+2i(\theta^{++\alpha}\theta^{0\beta})\partial^A_{\alpha\beta}
\bar f^iu^-_i\,.
\label{hyp-comp}
\eea
All component fields are defined on the $d{=}3$ Minkowski space $x_A^m\,$. With the auxiliary
fields being eliminated,
the superfield action \p{qFree} yields the following action for the physical fields:
\be
S_{phys}
= -\int d^3x(\bar f_i \square f^i
+\frac i2\bar \psi^{\alpha}_i
\partial_{\alpha\beta}\psi^{i\beta}).
\ee
Note that the presence of an infinite number of the auxiliary fields is an unavoidable feature of the formulation of the $d{=}3$
hypermultiplets with off-shell $\cN{=}3$ supersymmetry, in a full similarity to off-shell $\cN{=}2, d{=}4$ hypermultiplets.

When the superfield $q^+$ is placed in some representation of the gauge group,
\be
\delta q^+ = \Lambda q^+\,,
\ee
its minimal coupling to the gauge superfield $V^{++}$ is given by
\be S=\int d\zeta^{(-4)} \bar q^+({\cal D}^{++}
+V^{++})q^+.
\label{hyp}
\ee
At the moment we do not specify neither gauge group nor representation of the latter
on $q^+$; the specific cases we shall consider in the next sections correspond to some detailing
of the general gauged action \p{hyp}.

\subsection{$\cN{=}3$ superconformal transformations}
It is easy to construct the odd part of the $\cN{=}3$ superconformal transformations
of the coordinates of the initial $\cN{=}3$ superspace:
\bea
\delta_{sc} x^{\alpha\beta}&=&-i\epsilon^{kl\beta}\theta^\alpha_{kl}
-i\epsilon^{kl\alpha}\theta^\beta_{kl}\nn\\
&&-\frac{i}2\eta^{kl}_{\gamma}\theta^\alpha_{kl}x^{\gamma\beta}
-\frac{i}2\eta^{kl}_{\gamma}\theta^\beta_{kl}x^{\gamma\alpha}
+\frac12\eta^{kl}_{\rho}\theta^\alpha_{kl}\theta^{jn\rho}\theta^\beta_{jn}
+\frac12\eta^{kl}_{\rho}\theta^\beta_{kl}\theta^{jn\rho}\theta^\alpha_{jn}
,\nn\\
\delta_{sc}\theta^\alpha_{kl}&=&\epsilon^\alpha_{kl}+\frac12x^{\alpha\beta}\eta_{kl\beta}
-i\theta^\alpha_{jn}\theta^\gamma_{kl}\eta^{jn}_{\gamma}+\frac{i}2
\theta^{jn\alpha}\theta^\beta_{jn}\eta_{kl\beta},
\eea
where $\epsilon^\alpha_{kl}$ and $\eta^\alpha_{kl}$ are parameters of $Q$ and $S$
supersymmetries. All even superconformal transformations are contained in the Lie brackets of
these odd transformations. The full measure $d^3x d^6\theta$ is invariant under the $\cN{=}3$
superconformal group.

The superconformal transformations of the harmonics can be defined by analogy with
the $\cN{=}2, d{=}4$ case \cite{book},
\be
\delta_{sc}u^+_k=\lambda^{++}u^-_k,\qquad \delta_{sc}u^-_k=0\,,
\ee
where
\be
\lambda^{++}=-i\theta^{++\alpha}\theta^{0\beta}k_{\alpha\beta}-i\theta^{++\alpha}
u^+_ku^-_l\eta^{kl}_\alpha
+i\theta^{0\alpha}u^+_ku^+_l\eta^{kl}_\alpha
+u^+_ku^+_l\omega^{kl}\,.
\ee
Here $k_{\alpha\beta}$ and $\omega^{kl}$ are parameters of the special conformal
and $\textrm{SU}(2)_c$ transformations. The transformations of the analytic
$\cN{=}3$ coordinates under the $S$ supersymmetry and $\textrm{SU}(2)_c$ symmetry are
\bea
\delta_{sc} x^m_A&=&-i(\gamma^m)_{\alpha\beta} [\,x^{\beta\rho}_Au^-_ku^-_l\eta^{kl}_\rho
\theta^{++\alpha} - x^{\rho\beta}_Au^+_ku^-_l
\eta^{kl}_\rho\theta^{0\alpha} - 2\omega^{kl}u^-_k u^-_l\theta^{(\alpha\,++}\theta^{\beta) 0}\,]\,,
\nn\\
\delta_{sc}\theta^{0 \alpha}&=&
\frac12x^{\alpha\beta}_Au^+_k u^-_l\eta^{kl}_\beta
-iu^-_k u^-_l\eta^{kl}_\gamma\theta^{++ \alpha}\theta^{0 \gamma}
-\frac{i}2u^+_k u^-_l\eta^{kl\alpha}(\theta^0)^2 +\omega^{kl}u^-_k u^-_l \theta^{++ \alpha}\,, \nn\\
\delta_{sc}\theta^{++ \alpha}&=&
\frac12x^{\alpha\beta}_Au^+_k u^+_l\eta^{kl}_\beta
+\frac{i}2\eta^{kl\alpha}[u^-_k u^-_l(\theta^{++})^2
-u^+_k u^+_l(\theta^0)^2] + 2\omega^{kl}u^-_k u^+_l \theta^{++ \alpha}. \lb{N3conf1}
\eea

The transformations of the harmonic derivatives have the  form
\bea
&&\delta_{sc}\cD^{++}=-\lambda^{++}\cD^0,\quad
\delta_{sc}\cD^{--}=-(\cD^{--}\lambda^{++})\cD^{--}.
\eea
It is easy to find the superconformal transformation of the
analytic integration measure
\bea
\delta_{sc}d\zeta^{(-4)}=-2\lambda d\zeta^{(-4)},\quad \cD^{++}\lambda=\lambda^{++}\,.
\eea
Here
\be
\lambda=-\frac12d-\frac12x^{\alpha\beta}_A k_{\alpha\beta}
+i\theta^{0 \alpha}\eta^0_\alpha-i\theta^{++ \alpha}\eta^{--}_\alpha
+u^+_ku^-_l\omega^{kl},
\ee
$d$ being the scale transformation parameter.

The $\cN{=}3$ Chern-Simons action (\ref{CS}) and the minimal $q^+,V^{++}$ interaction
(\ref{hyp}) are invariant under the $\cN{=}3$ superconformal group realized on the basic
superfields as
\be
\delta_{sc}V^{++}=0,\quad \delta_{sc}q^+=\lambda q^+.
\ee
The $\cN{=}3$, $d{=}3$ action \p{3SYM} is obviously not superconformal because
of the presence of dimensionful coupling constant.

For the future use, it is worthwhile to point out that the
requirement of superconformal invariance forbids any
self-interaction of the hypermultiplets off shell: their only
superconformal off-shell actions are the free $q^+$ action
\p{qFree} and its minimal gauge covariantization
\p{hyp} \footnote{This uniqueness of superconformal $q^+$ action
can be understood also on the dimensionality grounds: the
analytic superspace integration measure has dimension $-1$ (in
mass units) while $[q^+] = 1/2$; so the action without dimensionful parameters
can be at most bilinear in $q^+$ superfields.}.

\setcounter{equation}{0}
\section{The ABJM model in $\cN{=}3$ harmonic superspace}

\subsection{Free hypermultiplets}
It is well known that the component content of the $\cN{=}6$ supersymmetric model
is given by four complex scalar fields and four complex spinor fields.
In the $\cN{=}3$ superfield formalism, these degrees of freedom can be described by two
hypermultiplet superfields $q^{+a}=\varepsilon^{ab}q^+_b$, $a,b=1,2$,
 and their conjugate
$\bar q^{+}_{a}=\widetilde{(q^{+a})}\,$, $\widetilde{( q^{+}_{b})}=- \bar{q}^{+b}\,$,
 with the action
\be
S_{free}= \int d\zeta^{(-4)} \bar q^{+}_a{\cal D}^{++}q^{+a}\,.
\label{hyp1}
\ee
This action is manifestly invariant under the extra $\textrm{SU}(2)_{ext}$ group acting
on the doublet indices $a$ and commuting with the $\cN{=}3$ supersymmetry. It also exhibits
an extra $\tU(1)$ symmetry realized as a common phase transformation of $q^{+a}$:
\be
q^{+ a}{}' = e^{i\tau} q^{+ a}\,, \quad \bar q^{+}_a{}' = e^{-i\tau} \bar q^{+}_a\,. \lb{U1}
\ee

\subsubsection{Extra supersymmetry}
The additional (hidden) supersymmetry transformations of the $\cN{=}3$ superfields
are defined through the spinor derivative $D^0_\alpha$ preserving
the Grassmann analyticity:
\be
\delta_\epsilon q^{+a}=i\epsilon^{\alpha\,ab} D^0_\alpha q^{+}_{b}
=-\widetilde{(\delta_\epsilon\bar q^{+}_{a})}\,,\quad
\delta_\epsilon\bar q^{+}_{a}=i\epsilon^{\alpha}_{ab} D^0_\alpha\bar
q^{+b}=\widetilde{(\delta_\epsilon q^{+a})}\,,
\label{epsilon}
\ee
where $\epsilon^{ab}_\alpha = \epsilon^{ba}_\alpha$ is a real spinor parameter, triplet of the
extra $\textrm{SU}(2)$ group, $\widetilde{(\epsilon^{ab}_\alpha)}=\epsilon_{\alpha\,ab}\,$.
Note the conjugation rule
$$
\widetilde{(D^0_\alpha q^{+}_{b})} =D^0_\alpha \bar{q}^{+b}\,.
$$
The free hypermultiplet action (\ref{hyp1}) is easily checked to be invariant under these
transformations
\be
\delta_\epsilon S_{free}= i\int d\zeta^{(-4)}
\epsilon^{\alpha\, ab}D^0_\alpha (\bar q^+_{a}{\cal D}^{++} q^+_{b})=0.
\ee
To show that (\ref{epsilon}) indeed generate supersymmetry, we
compute the commutator of two transformations (\ref{epsilon}) with
the spinor parameters $\epsilon_\alpha^{ab}$ and $\mu_\alpha^{ab}$,
\bea
[\delta_\mu\delta_\epsilon-\delta_\epsilon\delta_\mu]q^{+a}
&=&-\frac{1}{2}(\mu^{\alpha}{}^a_b\epsilon^{\beta\, bc}
+\mu^{\beta}{}^c_b\epsilon^{\alpha\,ab})
\left[\{D^0_\alpha, D^0_\beta \}- \varepsilon_{\alpha\beta}
(D^0)^2\right]q^{+}_{c}\nn\\
&=&\frac i2\mu^{(\alpha}_{bc}\epsilon^{\beta)\,bc}
\partial^A_{\alpha\beta}q^{+a}
-\mu^{\alpha}{}^{(a}_{\ b}\epsilon_{\alpha}^{c)b}
(D^0)^2 q^{+}_{c}.
\label{131}
\eea
The last term in (\ref{131}) vanishes on shell,
$D^{++}q^+_a=0\ \Rightarrow\ (D^0)^2q^+_a=0$. As a result,
the commutator of two transformations (\ref{epsilon})
generates the $x$-translations of hypermultiplets with the bosonic
parameter $\mu^{(\alpha}_{bc}\epsilon^{\beta)\,bc}$ and, hence, (\ref{epsilon})
do form three supersymmetries on shell.
These three additional supersymmetry transformations,
together with three explicit $\cN{=}3$ ones, constitute
the $\cN{=}6$ invariance of the free hypermultiplet action (\ref{hyp1}). Note that the Lie bracket of the implicit
and explicit supersymmetry transformations is vanishing as a consequence of the anticommutativity of $D^0_\alpha$
and the $\cN{=}3$ supersymmetry generators.

\subsubsection{$\textrm{SO}(6)$ R-symmetry}
The free action of two hypermultiplet superfields also exhibits an
invariance under the full automorphism group $\textrm{SO}(6)$ of the $\cN{=}6$
superalgebra.

The action (\ref{hyp1}) is explicitly
invariant only under the group $\textrm{SU}(2)_R\times \textrm{SU}(2)_{ext}$, where
$\textrm{SU}(2)_R$ is the group of internal automorphisms of $\cN{=}3$ harmonic
superspace while $\textrm{SU}(2)_{ext}$ is realized on the index $a$ in this
action. Therefore, to show the invariance of the action under the
full $\textrm{SO}(6)$ R-symmetry group we need to specify the
remaining transformations from the coset $\textrm{SO}(6)/[\textrm{SU}(2)_R\times \textrm{SU}(2)_{ext}]$.
This coset is parametrized by nine real parameters,
\be
\lambda^{(ij)(ab)},
 \quad \overline{(\lambda^{(ij)(ab)})}
 = \lambda_{(ij)(ab)}.
\label{1}
\ee
The linear realization of these transformations on the physical
scalar fields $f^{i\, a}$ can be chosen as
\be
\delta_\lambda f^{i\,a} = i \lambda^{(ij)(ab)}
 f_{j\,b}, \quad
\delta_\lambda \bar f_{i\,a}
= -i \lambda_{(ij)(ab)}\bar f^{j\,b}, \quad
\bar f_{i\,a}= \overline{ f^{i\,a}},
\label{lam}
\ee
so that $f^{i\,a}\bar f_{i\,a}$ is the full $\textrm{SO}(6)$ invariant.
These physical scalars appear in the lowest order of the component
expansion of the hypermultiplets,
$q^{+a}=u^+_i f^{i\,a}+\ldots$,
$\bar q^+_a=-u^+_i \bar f^i_a +\ldots$. Therefore there should
be a generalization of the transformations (\ref{lam}) for the
hypermultiplet superfields.

This generalization is unambiguously determined by requiring the variation $\delta_\lambda q^+$ to have
the same harmonic $\tU(1)$ charge $+1$ as $q^+$ itself and to be analytic.
We project the parameters $\lambda^{(ij)(ab)}$ on the harmonic variables,
\be
\lambda^{\pm\pm(ab)}=u^\pm_i u^\pm_j \lambda^{(ij)(ab)},\qquad
\lambda^{0(ab)}=u^+_i u^-_j \lambda^{(ij)(ab)},
\ee
and define the hidden $\textrm{SO}(6)$ transformation of the hypermultiplet superfields as
\bea
\delta_\lambda q^{+ a}&=& -i[ \lambda^{0(ab)}- \lambda^{++(ab)}\hat{\cal D}^{--}
- 2\lambda^{--(ab)}\theta^{++ \alpha}D^0_\alpha
+ 4\lambda^{0(ab)}\theta^{0 \alpha}D^0_\alpha]q^+_{b},\nn\\
\delta_\lambda \bar{q}^{+}_a&=& -i[ \lambda^{0}_{(ab)}- \lambda^{++}_{(ab)}\hat{\cal D}^{--}
- 2\lambda^{--}_{(ab)}\theta^{++ \alpha}D^0_\alpha
+ 4\lambda^{0}_{(ab)}\theta^{0 \alpha}D^0_\alpha]\bar q^{+ b}. \label{44}
\eea
Here $\hat{\cal D}^{--}$ is a modification of the harmonic
derivative ${\cal D}^{--}$ such that $\hat{\cal D}^{--}$ preserves
analyticity,
\be
\hat {\cal D}^{--} ={\cal D}^{--} + 2\theta^{--\alpha}D^0_\alpha
= \partial^{--} + 2\theta^{0 \alpha}\frac{\partial}{\partial \theta^{++ \alpha}}\,,
\quad [D^{++}_\alpha,\hat{\cal D}^{--}] = 0. \label{5}
\ee
One can easily check that under the superfield transformations (\ref{44}) the lowest bosonic components of the
hypermultiplet superfields transform as is (\ref{lam}) while the transformations of
the auxiliary fields coming from the harmonic expansions are not essential here
since these fields vanish on shell.

With the help of the following identity
\be
{\cal D}^{++}\delta_\lambda q^{+a}
=-i[ \lambda^{0(ab)}
- \lambda^{++(ab)}\hat{\cal D}^{--}
- 2\lambda^{--(ab)}\theta^{++\alpha}D^0_\alpha
+ 4\lambda^{0(ab)}\theta^{0 \alpha}D^0_\alpha]
{\cal D}^{++} q^+_{b},
\ee
we compute the variation of the action
(\ref{hyp1}),
\be
\delta_\lambda S_{free}=
-i\int d\zeta^{(-4)}[2\lambda^0_{(ab)}\bar q^{+a}{\cal D}^{++}q^{+b}
 -\lambda^{++}_{(ab)}\hat{\cal D}^{--}(\bar q^{+a}{\cal D}^{++}q^{+b})
 +4\lambda^{0}_{(ab)}\theta^{0\alpha}D^0_\alpha (\bar
 q^{+a}{\cal D}^{++}q^{+b})].
\label{72}
\ee
The last term in (\ref{72}) is a total derivative, while after
integration by parts the second term cancels the first one.
Thus the free hypermultiplet action \p{hyp1} is invariant under (\ref{44}),
\be
\delta_\lambda S_{free}=0.
\ee

Due to the presence of explicit $\theta$s in the transformation \p{44}, it does not commute with
the manifest $\cN{=}3$ supersymmetry. It is easy to show that, modulo equations of motion for $q^{+a}$, this commutator
yields just the hidden $\cN{=}3$ supersymmetry \p{epsilon}. We shall discuss this closure in more detail later on,
in the non-trivial interaction cases. It is worth noting that the closure of the hidden
$\textrm{SO}(6)$ transformations \p{44} (and their generalization to the interaction case) contains $\textrm{SU}(2)_{ext}$ and
just the superconformal R-symmetry group $\textrm{SU}(2)_c$
defined in \p{N3conf1}. The latter becomes indistinguishable from the standard $\textrm{SU}(2)_R$ only after elimination of the
hypermultiplet auxiliary fields by their equations of motion, i.e.\ on shell.
Note also that the $\tU(1)$ symmetry \p{U1} commutes with both hidden and manifest
$\cN{=}3$ supersymmetries (as well as with the extra $\textrm{SO}(6)$ transformations).

In fact, the symmetry of the action (\ref{hyp1}) is even wider than $\cN{=}6$ supersymmetry plus $\textrm{SO}(6)$ R-symmetry:
it is the maximal $\cN{=}8$ on-shell supersymmetry
together with its automorphism symmetry $\textrm{SO}(8)$. We postpone discussion of these additional
symmetries until Sect.\ 5, where they will be considered at the full interaction level. The off-shell superconformal $\cN{=}3$
invariance of \p{hyp1} taken together with its on-shell $\textrm{SO}(8)$ R-symmetry and $\cN{=}8$ supersymmetry imply that the free action
of two $d{=}3$ hypermultiplets on shell (i.e.\ modulo algebraic equations of motion for the auxiliary fields) respects
the maximal $\cN{=}8, d{=}3$ superconformal symmetry.

\subsection{The $\tU(1)\times \tU(1)$ theory}
As the next step, we consider the $\tU(1)\times \tU(1)$ gauge theory. This simplest example with interaction
will be used to further explain the basic ideas of  our construction.

\subsubsection{Actions}
Now we have two Abelian gauge superfields $V^{++}_L$ and
$V^{++}_R$ corresponding to the two $\tU(1)$ gauge groups. In accord
with the proposal of \cite{ABJM}, the gauge action for these
superfields should be a difference of two Chern-Simons
actions (\ref{CS}). In the Abelian case, such action is very simple:
\bea
S_{gauge} &=& S_{CS}[V^{++}_L]-S_{CS}[V^{++}_R] \nn \\
&=& -\frac{ik}{8\pi}\int d^9z du_1 du_2\frac{1}{(u^+_1u^+_2)^2}\left[
V^{++}_L(z,u_1)V^{++}_L(z,u_2) - V^{++}_R(z,u_1)V^{++}_R(z,u_2)\right] \nn \\
&=& -\frac{ik}{8\pi} \int d\zeta^{(-4)}\left(V^{++}_L W^{++}_L - V^{++}_R W^{++}_R \right),
\label{gauge2}
\eea
where we used the relation \p{MeaS} and the definition \p{W}.
The gauge invariant generalization of the hypermultiplet action
(\ref{hyp1}) is
\be
S_{hyp}= \int d\zeta^{(-4)} \bar q^{+}_{a}({\cal D}^{++}+V^{++}_L-V^{++}_R)q^{+a}
= \int d\zeta^{(-4)} \bar q^{+}_{a}\nabla^{++}q^{+a}.
\label{hyp2}
\ee
Note that the gauge covariant harmonic derivative $\nabla^{++}=
{\cal D}^{++}+V^{++}_L-V^{++}_R$ depends only on the difference of
two gauge superfields, but not on their sum (cf.\ the corresponding
covariant space-time derivatives given in \cite{ABJM,BLS1}). So it is useful
to define new gauge superfields,
\be
V^{++}_L+V^{++}_R = V^{++}\,, \quad  V^{++}_L-V^{++}_R = A^{++}\,,
\ee
in terms of which \p{gauge2} and \p{hyp2} are rewritten as
\bea
S_{gauge} &=&  -\frac{ik}{8\pi} \int d\zeta^{(-4)} \,V^{++} W^{++}_{(A)} =
-\frac{ik}{8\pi} \int d\zeta^{(-4)} \,A^{++} W^{++}_{(V)}\,, \lb{VA} \\
S_{hyp}&=& \int d\zeta^{(-4)} \bar q^{+}_{a}({\cal D}^{++}+ A^{++})q^{+a}\,.\lb{hyp22}
\eea
The action \p{hyp22} is invariant under the following gauge transformations
\be
q^{+ a}{}' = e^{\Lambda} q^{+ a}\,, \; \bar q^{+}_{a}{}' = e^{-\Lambda} \bar q^{+}_{a}\,, \;A^{++}{}' = A^{++} - {\cal D}^{++}\Lambda\,, \qquad
\Lambda = \Lambda_L - \Lambda_R\,. \lb{u1u1HypG}
\ee
The rest of the gauge group $\tU(1)\times \tU(1)$, with the gauge parameter $\hat{\Lambda} = \Lambda_L + \Lambda_R$, acts only on $V^{++}$ and
does not affect hypermultiplets at all.

In the considered case, the general variation formula \p{var-CS} is written as
\bea
\delta S_{gauge} &=& -\frac{ik}{4\pi} \int d\zeta^{(-4)} \,\left(\delta V^{++}_L W^{++}_L - \delta V^{++}_R W^{++}_R\right) \nn \\
&=& -\,\frac{ik}{8\pi} \int d\zeta^{(-4)} \,\left(\delta V^{++} W^{++}_{(A)} + \delta A^{++} W^{++}_{(V)}\right)\,.\lb{var-CS2}
\eea
It is also instructive to present the full set of superfield equations for the $\tU(1)\times \tU(1)$
case:
\be
(\mbox{a}) \;\nabla^{++}q^{+a} = \nabla^{++}\bar q^{+a} = 0;
\quad (\mbox{b}) \; W^{++}_L  = W^{++}_R = -i\frac{4\pi}{k}\bar q^+_a q^{+ a}\,.\lb{EqMo}
\ee

The $\tU(1)\times \tU(1)$ Chern-Simons and hypermultiplet actions are invariant
under the $P$-parity transformation
\bea
&&Px^{0,2}_A=x^{0,2}_A,\quad Px^1_A=-x^1_A,\quad P\theta^{0,\pm\pm}_\alpha=
-(\gamma_1)_\alpha^\beta\theta^{0,\pm\pm}_\beta,\quad
P(\theta^{0,\pm\pm})^2=-(\theta^{0,\pm\pm})^2,\nn\\
&&PD^{0,++}_\alpha=(\gamma_1)_\alpha^\beta D^{0,++}_\beta,\quad
P(D^{0,++})^2=-(D^{0,++})^2. \eea
The parity of the superfields can
be chosen as follows
\bea
&&P V^{\pm\pm}_L(\zeta_P)=V^{\pm\pm}_R(\zeta),\quad
PW^{++}_L(\zeta_P)=-W^{++}_R(\zeta),\nn\\
&&Pq^{+a}(\zeta_P)=\bar{q}^{+a}(\zeta),\quad P \bar{q}^+_a(\zeta_P)=q^+_a(\zeta).
 \eea

\subsubsection{$\cN{=}6$ supersymmetry}
Now we are going to prove that the sum of the gauge and
matter actions (\ref{gauge2}), (\ref{hyp2}),
\be
S_{\cN{=}6}=S_{gauge}+S_{hyp}\,,
\label{S6-ab}
\ee
possesses the $\cN{=}6$ supersymmetry. To this end, as the first step, we generalize the transformations
(\ref{epsilon}),
\bea
\delta_\epsilon q^{+a}&=&i\epsilon^{\alpha\,(ab)}
 [\nabla^0_\alpha +\theta^{--}_\alpha(W^{++}_L-W^{++}_R)]
 q^{+}_{b},\nn\\
\delta_\epsilon\bar q^{+}_{a}&=&i\epsilon^{\alpha}_{(ab)}
 [\nabla^0_\alpha -\theta^{--}_\alpha(W^{++}_L-W^{++}_R)]\bar
 q^{+b}.
\label{epsilon1}
\eea
Here $\nabla^0_\alpha$ is a gauge-covariant generalization of the
spinor derivative $D^0_\alpha\,$. It acts on the hypermultiplets
according to
\be
\nabla^0_\alpha q^+_a
=(D^0_\alpha+V^0_{L\,\alpha}-V^0_{R\,\alpha})q^+_a,\qquad
\nabla^0_\alpha \bar q^{+a}
=(D^0_\alpha-V^0_{L\,\alpha}+V^0_{R\,\alpha})\bar q^{+a}.
\ee
The gauge potentials $V^0_{L,R\,\alpha}$ are expressed through
$V^{--}_{L,R}$ as $V^0_{L,R\,\alpha}=-\frac12D^{++}_\alpha
V^{--}_{L,R}$, where $V^{--}_{L}=V^{--}_{L}(V^{++}_L)$ and
$V^{--}_{R}=V^{--}_{R}(V^{++}_R)$ appear as the solutions of
the Abelian version of zero-curvature equation (\ref{zero-curv}).
In contrast to the flat spinor derivative $D^0_\alpha$,
the gauge-covariant derivative $\nabla^0_\alpha$ does not preserve
the analyticity,
\be
[D^{++}_\alpha,\nabla^0_\beta]=
-\frac14\varepsilon_{\alpha\beta}(D^{++})^2(V^{--}_L-V^{--}_R)
=\varepsilon_{\alpha\beta}(W^{++}_L-W^{++}_R).
\ee
However, one can check that the expression $\nabla^0_\alpha
+\theta^{--}_\alpha(W^{++}_L-W^{++}_R)$ entering the
transformations (\ref{epsilon1}) does preserve analyticity,
\be
[D^{++}_\alpha,\nabla^0_\beta
+\theta^{--}_\beta(W^{++}_L-W^{++}_R)]=0\,.
\ee

Now we compute the variation of the hypermultiplet action
(\ref{hyp2}) with respect to the transformation (\ref{epsilon1}),
\bea
\delta_\epsilon S_{hyp}&=&i\int d\zeta^{(-4)}
[\epsilon^{\alpha\,ab}D^0_\alpha(\bar q^+_a\nabla^{++} q^+_b)
+2\epsilon^{\alpha\,ab}\theta^0_\alpha (W^{++}_L-W^{++}_R)\bar q^+_a
q^+_b]\nn\\
&=&2i\int d\zeta^{(-4)} \epsilon^{\alpha\,ab}\theta^0_\alpha
(W^{++}_L-W^{++}_R)\bar q^+_a q^+_b.
\label{43}
\eea
The non-vanishing expression in the second line of (\ref{43}) can
be compensated by the following transformation of the gauge
superfields,
\be
\delta V^{++}_L=\delta V^{++}_R= \frac{8\pi}k\epsilon^{\alpha\,ab}
\theta^0_\alpha \bar q^+_a q^+_b.
\label{delta-V}
\ee
Indeed, applying the formula (\ref{var-CS2}) for the variation of the Chern-Simons action, we find
\bea
\delta_\epsilon S_{gauge}
=-2i\int d\zeta^{(-4)}\epsilon^{\alpha\,ab}\theta^0_\alpha
\bar q^+_a q^+_b(W^{++}_L-W^{++}_R),
\eea
which exactly cancels \p{43}. Note that the gauge superfield $A^{++} = V^{++}_L-V^{++}_R$,
which appears in the hypermultiplet action (\ref{hyp22}), is inert
under the transformations (\ref{delta-V}), $\delta_\epsilon A^{++} = 0\,$.
Thus we conclude that the total action (\ref{S6-ab}) is invariant under the three extra
supersymmetry transformations realized on the involved $\cN{=}3$ superfields
by the rules (\ref{epsilon1}), (\ref{delta-V}).

The last issue is to show that the commutator
of two consequent transformations (\ref{epsilon1}) generates on shell $x^m$-translations of
the superfields,
\bea
[\delta_\mu\delta_\epsilon-\delta_\epsilon\delta_\mu]q^{+a}
&=&-\frac{1}{2}(\mu^{\alpha}{}^a_b\epsilon^{\beta\, bc}
-\mu^{\beta}{}^c_b\epsilon^{\alpha\,ab})
\left[\{\nabla^0_\alpha, \nabla^0_\beta \}
-\varepsilon_{\alpha\beta}
(\nabla^0)^2\right]q^{+}_{c}\nn\\
&=& \frac i2\mu^{(\alpha}_{bc}\epsilon^{\beta)\,bc}
\nabla_{\alpha\beta}q^{+a}
-\mu^{\alpha}{}^{(a}_{\ b}\epsilon_{\alpha}^{c)b}
(\nabla^0)^2 q^{+}_{c}\,.
\label{136}
\eea
Here $\nabla_{\alpha\beta}$
is a gauge covariant $d{=}3$ vector derivative
which generates the ``covariant'' translations with the bosonic parameter
$\mu^{(\alpha}_{bc}\epsilon^{\beta)\,bc}$ (it is a sum of ordinary $x$-translation and a field-dependent $\tU(1)$ gauge transformation).
The last term in
(\ref{136}) vanishes on shell. To show this, we first note that, by analyzing the harmonic differential equations,
one can prove (see \cite{book} for the details)
\be
\nabla^{++} q^+_a=0 \quad\Rightarrow\quad \nabla^{--}\nabla^{--}
q^+_a=0.
\ee
Next, using the analyticity of $q^+_a$ we have
\be
0=D^{++\alpha}D^{++}_\alpha\nabla^{--}\nabla^{--} q^+_a
=-8[(W^{++}_L-W^{++}_R)\nabla^{--}+ (W_L^0-W^0_R)
-\nabla^{0\alpha}\nabla^0_\alpha]q^+_a,
\label{48}
\ee
where $W^0_{L,R}=\frac12 {\cal D}^{--}W^{++}_{L,R}$.
Now we exploit the equations of motion for the gauge superfields, (\ref{EqMo}b),
to deduce their corollaries
\be
W^{++}_L-W^{++}_R=0,\qquad
W^0_L-W^0_R=0\,.
\label{50}
\ee
Then (\ref{48}) implies
\be
\nabla^{0\alpha}\nabla^0_\alpha q^+_a=0\,.
\ee
This completes the proof that on shell the commutator \p{136} of two
extra $\cN{=}3$ supersymmetry transformations (\ref{epsilon1}) yields, modulo a field-dependent gauge
transformation,  the ordinary $d{=}3$ translation.

The $\tU(1)\times \tU(1)$ model also respects the appropriate generalization of the free case $\textrm{SO}(6)$ R-symmetry \p{44}.
We shall postpone discussion of this symmetry until considering the gauge group $\tU(N)\times \tU(M)$ in the
next Subsection. The $\tU(1)\times \tU(1)$ example follows from this more general case via an obvious reduction.

\subsection{The $\tU(N)\times \tU(M)$ theory}

The crucial idea in constructing the
$\cN{=}6$ supersymmetric gauge theory in \cite{ABJM}
 was to consider the matter
fields in the bifundamental representation of the $\tU(N)\times
\tU(N)$ gauge group, the product of the fundamental representation of the left $\tU(N)$
and the conjugated fundamental representation of the right $\tU(N)$. Actually, one can consider the more general case
of the gauge group $\tU(N)\times \tU(M)\,, \;N\neq M$ (see, e.g., \cite{Hosomichi,STac,ABJ,Li}):
\be
(N,\bar{M}):\quad (q^{+a})_A^{\underline B},\qquad (\bar{N},M):\quad(\bar{q}^+_a)^A_{\underline B}\,,\label{bi1}
\ee
where $A=1,\ldots, N$ and $\underline{B}=1,\ldots, M\,$. Hereafter, the underlined indices refer
to the right $\tU(M)$ gauge group. Yet admissible is another type of the bifundamental representation,
the product of two fundamental representations \cite{Li}:
\bea
(N,M):\quad (q^{+a})_{A\underline{B}},\quad (\bar{N},\bar{M}):
\quad(\bar{q}^+_a)^{A\underline{B}}\,.
\label{bi2}
\eea
In this Subsection we shall focus on the case \p{bi1} as the standard and most instructive one.
The case \p{bi2} as well as some other options admitting hidden supersymmetries  will be shortly
addressed in the Section 4.

The gauge superfields for the groups $\tU(N)$ and $\tU(M)$ are
given by the antihermitian matrices $(V^{++}_L)^A_B$ and
$(V^{++}_R)^{\underline{A}}_{\underline{B}}$. The gauge
interaction of the hypermultiplets with the gauge superfields in
the $(N,\bar{M})$-model under consideration reads
\bea
&&(\nabla^{++}q^{+a})_A^{\underline{B}}={\cal D}^{++}(q^{+a})_A^{\underline{B}}
 + (V^{++}_L)_A^B (q^{+a})_B^{\underline{B}}
 - (q^{+a})_A^{\underline{A}}(V^{++}_R)_{\underline{A}}^{\underline{B}}\,, \nn \\
&& (\nabla^{++}\bar q^{+}_a)^A_{\underline{B}}={\cal D}^{++}(\bar q^{+}_a)^A_{\underline{B}}
 - (\bar q^{+}_a)^B_{\underline{B}}(V^{++}_L)^A_B
 + (V^{++}_R)^{\underline{A}}_{\underline{B}}(\bar q^{+}_a)^A_{\underline{A}}\,. \lb{GaugeNabl}
 \eea
The matrix form of the $(N,\bar{M})$ harmonic derivative is
\bea
&&\nabla^{++}q^{+a}={\cal D}^{++}q^{+a}
 + V^{++}_L q^{+a}- q^{+a} V^{++}_R\,, \;  \nabla^{++}\bar q^{+}_a={\cal D}^{++}\bar q^{+}_a
 -\bar q^{+}_a V^{++}_L +  V^{++}_R \bar q^{+}_a\,, \nn\\
&&\bar{q}^+_a=(q^{+a})^\dagger.
\eea
The superfield $\nabla^{++}q^{+a}$ transforms covariantly under the
following infinitesimal gauge transformations
\bea
\delta q^{+a} &=& \Lambda_L q^{+a}-q^{+a}\Lambda_R,\qquad
 \delta\bar q^{+}_{a} = \Lambda_R \bar q^{+}_{a}-\bar q^{+}_{a}\Lambda_L,\nn\\
\delta V^{++}_L&=&-{\cal D}^{++}\Lambda_L -[V^{++}_L,\Lambda_L],\qquad
\delta V^{++}_R=-{\cal D}^{++}\Lambda_R -[V^{++}_R,\Lambda_R].
\eea
The analytic gauge parameters $\Lambda_L$ and $\Lambda_R$ are antihermitian matrices,
$\Lambda_L^\dag=-\Lambda_L$,
$\Lambda_R^\dag=-\Lambda_R$.
The hermitian conjugation for the other superfields is defined as
\be
(q^{+a})^\dag= \bar q^{+}_{a},\quad
(\bar q^{+}_{a})^\dag = -q^{+a},\quad
(V^{++}_L)^\dag=-V^{++}_L,\quad
(V^{++}_R)^\dag=-V^{++}_R.
\ee

After fixing the notations, we turn to the actions.
We write down the non-Abelian $\cN{=}6$ supersymmetric
action as a direct generalization of the $\tU(1)\times \tU(1)$ actions \p{S6-ab},
\p{gauge2}, \p{hyp2}:
\bea
S_{\cN{=}6}&=&S_{gauge}+S_{hyp},\label{N6}\\
S_{gauge}&=&S_{CS}[V^{++}_L]-S_{CS}[V^{++}_R],\label{Sgauge}\\
S_{hyp}&=&\tr\int d\zeta^{(-4)}\bar q^{+}_{a}\nabla^{++} q^{+a},
\label{hyp3}
\eea
where the Chern-Simons action $S_{CS}[V^{++}]$ is given by
(\ref{CS}). The analytic superfield equations of motion corresponding to \p{N6} read
\bea
&& (\nabla^{++} q^{+a})^{\underline{A}}_B = (\nabla^{++} \bar q^{+a})_{\underline{A}}^B  = 0\,,\lb{Qeq} \\
&& (W^{++}_L)_A^B  = -i\frac{4\pi}{k}\,(q^{+ a})_A^{\underline{D}}(\bar q^{+}_{a})^B_{\underline{D}}\,, \quad
(W^{++}_R)_{\underline{A}}^{\underline{B}} = -i\frac{4\pi}{k}\,(\bar q^{+}_{a})^D_{\underline{A}}(q^{+ a})_D^{\underline{B}}\,. \lb{VEqs}
\eea

\subsubsection{$\cN{=}6$ supersymmetry}
We claim that the action (\ref{N6}) is invariant under the
following three extra supersymmetry transformations:
\bea
\delta_\epsilon q^{+a}&=& i\epsilon^{\alpha\,(ab)}\hat\nabla^0_\alpha q^{+}_{b}\,, \quad
\delta_\epsilon\bar q^{+}_{a} = i\epsilon^{\alpha}_{(ab)} \hat\nabla^0_\alpha\bar q^{+ b}\,, \nn \\
\delta_\epsilon V^{++}_L&=&\frac{8\pi}k\epsilon^{\alpha\,(ab)}
 \theta^0_\alpha q^+_a\bar q^+_b\,, \quad
\delta_\epsilon V^{++}_R = \frac{8\pi}k\epsilon^{\alpha\,(ab)}
 \theta^0_\alpha \bar q^+_a q^+_b\,,
\label{epsilon4}
\eea
where
\bea
&& \hat\nabla^0_\alpha q^{+}_{b} = \nabla^0_\alpha q^{+}_{b} +\theta^{--}_\alpha (W^{++}_L q^{+}_{b} -q^{+}_{b} W^{++}_R )\,, \lb{nablahat}\\
&& \nabla^0_\alpha q^+_a=D^0_\alpha q^+_a
 +V^0_{L\,\alpha}q^+_a -q^+_aV^0_{R\,\alpha}\,, \quad V^0_{L,R\,\alpha}=-\frac12D^{++}_\alpha
V^{--}_{L,R} \nonumber
\eea
and $\hat\nabla^0_\alpha\bar q^{+ b}\,, \nabla^0_\alpha\bar q^{+ b}$ are obtained via the $\widetilde{\;\;}$ conjugation.
The modified gauge-covariant derivative $\hat\nabla^0_\alpha$ preserves the $\cN{=}3$
analyticity as opposed to  $\nabla^0_\alpha\,$.

The variation of the hypermultiplet action (\ref{hyp3})
under the transformations (\ref{epsilon4}) is
\bea
\delta_\epsilon S_{hyp}&=&2i\,\tr\int d\zeta^{(-4)}
\epsilon^{\alpha\, (ab)}\theta^0_\alpha
 (W^{++}_L q^+_a\bar q^+_b- W^{++}_R \bar
q^+_a q^+_b)\nn\\
&&+\frac{8\pi}k\,\tr\int d\zeta^{(-4)}\epsilon^{\alpha\, (ab)}\theta^0_\alpha
[q^+_a\bar q^+_b q^{+c}\bar q^+_c
+\bar q^+_{a} q^+_b \bar q^{+c} q^+_c].
\label{var-hyp1}
\eea
Using a simple Fierz rearrangement, one can check that
\be
\tr\left[ q^+_{(a}\bar q^+_{b)} q^{+c}\bar q^+_c
+\bar q^+_{(a} q^+_{b)} \bar q^{+c} q^+_c\right] =0\,,\lb{BasId}
\ee
so the variation (\ref{var-hyp1}) is reduced to
\be
\delta_\epsilon S_{hyp}= 2i\,\tr\int d\zeta^{(-4)}
\epsilon^{\alpha\, ab}\theta^0_\alpha
 (W^{++}_L q^+_a\bar q^+_b- W^{++}_R \bar
q^+_a q^+_b).
\label{var-hyp2}
\ee
This expression is exactly canceled by the variation of
the Chern-Simons term (\ref{Sgauge}),
\be
\delta_\epsilon S_{gauge}=-2i\,\tr\int d\zeta^{(-4)}
\epsilon^{\alpha\, ab}\theta^0_\alpha
(W^{++}_L q^+_a\bar q^+_b- W^{++}_R \bar q^+_a q^+_b).
\ee
As a result, we proved that the total action (\ref{N6}) is
invariant under the $\cN{=}3$ supersymmetry transformations
(\ref{epsilon4}),
\be
\delta_\epsilon S_{\cN{=}6}=\delta_\epsilon(S_{gauge}+S_{hyper}) =0.
\ee
Together with the manifest $\cN{=}3$ supersymmetries of the $\cN{=}3$
superspace, the transformations (\ref{epsilon4}) form $\cN{=}6$ supersymmetry.
Therefore we conclude that the action (\ref{N6}) provides the
formulation of the $(N,\bar{M})$ $\cN{=}6$ Chern-Simons model
in the $\cN{=}3$ harmonic superspace. Like in the $\tU(1)\times \tU(1)$ model,
the extra hidden $\cN{=}3$ supersymmetry, as opposed to the manifest $\cN{=}3$ one,
has the correct closure on $d{=}3$ translations only modulo the superfield equations of motion \p{Qeq}, \p{VEqs}
and a field-dependent gauge transformation, i.e.\ it is essentially on-shell.

\subsubsection{$\textrm{SO}(6)$ R-symmetry}
A natural generalization of the transformations (\ref{44})
to the $\tU(N)\times \tU(M)$ case is
\bea
\delta_\lambda q^{+ a}&=& -i[ \lambda^{0(ab)}- \lambda^{++(ab)}\hat\nabla^{--}
- 2\lambda^{--(ab)}\theta^{++ \alpha}\hat\nabla^0_\alpha
+ 4\lambda^{0(ab)}\theta^{0 \alpha}\hat\nabla^0_\alpha]q^+_{b},\nn\\
\delta_\lambda \bar{q}^{+}_a&=& -i[ \lambda^{0}_{(ab)}
- \lambda^{++}_{(ab)}\hat\nabla^{--}
- 2\lambda^{--}_{(ab)}\theta^{++ \alpha}\hat\nabla^0_\alpha
+ 4\lambda^{0}_{(ab)}\theta^{0 \alpha}\hat\nabla^0_\alpha]
\bar q^{+ b}, \label{non-ab}
\eea
where $\hat\nabla^{--}$ and $\hat\nabla^0_\alpha$ are gauge-covariant
analyticity-preserving derivatives:
\bea
\hat{\nabla}^0_\alpha &=& \nabla^0_\alpha + \theta^{--}_\alpha W^{++},\qquad
\{D^{++}_\alpha, \hat{\nabla}{}^0_\beta\} =0,\nn\\
\hat\nabla^{--} &=& \nabla^{--} + 2\theta^{\alpha--}\nabla^0_\alpha
 + (\theta^{--})^2 W^{++},\qquad
 [D^{++}_\alpha, \hat\nabla^{--}] = 0. \label{b}
\eea

The variation of the hypermultiplet action (\ref{hyp3}) under (\ref{non-ab}) is
\be
\delta_\lambda S_{hyp}= i\,\tr\int d\zeta^{(-4)}\kappa_{(ab)}
\bar q^{+a}(W^{++}_Lq^{+b}-q^{+b}W^{++}_R),
\label{delta-hyp}
\ee
where
\be
\kappa_{(ab)}=
4\lambda^{--}_{(ab)}(\theta^0\theta^{++})
- 8 \lambda^{0}_{(ab)}(\theta^0)^2 .
\ee
Here we have used the following identities
\be
[\nabla^{++}, \hat{\nabla}^0_\alpha]q^+_a
= 2\theta^0_\alpha (W^{++}_L q^+_a-q^+_a W^{++}_R), \qquad
[\nabla^{++},\hat\nabla^{--}]q^+_a = (1
+ 4 \theta^{0 \alpha}\hat{\nabla}^0_\alpha)q^+_a.
\ee
To cancel the variation (\ref{delta-hyp}) we have to make the
following transformation of the gauge superfields
\be
\delta_\lambda V^{++}_L = \frac{4\pi}k\kappa^{ab}
 q^+_{a}\bar q^+_{b}, \qquad
\delta_\lambda V^{++}_R= \frac{4\pi}k\kappa^{ab}
 \bar q^+_{a} q^+_{b},
 \label{var12}
\ee
under which the Chern-Simons action (\ref{Sgauge}) varies as
\bea
\delta_\lambda S_{gauge}&=&
-i\,\tr\int d\zeta^{(-4)}\kappa^{ab}(
 q^+_{a}\bar q^+_{b}W^{++}_L
 -\bar q^+_{a}q^+_{b} W^{++}_R).
\eea
The variations \p{var12} performed in the hypermultiplet action produce quartic $q^+$ terms
which cancel each other as a consequence of the same identity \p{BasId} as in the case of hidden $\cN{=}3$ supersymmetry.

As a result, we proved that the action (\ref{N6}) is invariant
under the transformations (\ref{non-ab}),
\be
\delta_\lambda S_{\cN{=}6}=0,
\ee
and, therefore, respects the $\textrm{SO}(6)$ R-symmetry group.

It is interesting to calculate the Lie bracket of the $\textrm{SO}(6)/\textrm{SO}(4)$ transformations with the manifest ${\cal N}=3$ supersymmetry.
In the analytic basis, the latter is realized by the following differential operator
\be
\delta_\epsilon = \epsilon^{0\alpha}\left(\frac{\partial}{\partial \theta^{0\alpha}}
+ 2i\theta^{0\beta}\partial_{\beta\alpha}\right)
+ \epsilon^{++\alpha}\,\frac{\partial}{\partial \theta^{++\alpha}}
+ \epsilon^{--\alpha}\left(\frac{\partial}{\partial \theta^{--\alpha}}
- 2i\theta^{++ \beta}\partial_{\beta\alpha}\right).\lb{Qq}
\ee
Then
\bea
[\delta_\lambda\delta_\epsilon - \delta_\epsilon\delta_\lambda]\,(V^{++}_L)^A_B =
\frac{8\pi}{k}\omega^\alpha_{(ab)}\,\theta^0_\alpha\;(q^{+ a})^{\underline{B}}_{B}\,(\bar q^{+ b})^{A}_{\underline{B}} +
\frac{4\pi}{k}(D^{++} f^{--}_{(ab)})\;(q^{+ a})^{\underline{B}}_{B}\,(\bar q^{+ b})^{A}_{\underline{B}}\,,  \lb{lie1}
\eea
\bea
[\delta_\lambda\delta_\epsilon - \delta_\epsilon\delta_\lambda](q^{+a})^{\underline{B}}_{A} &=&
i\omega^{\alpha (ab)}\left(\hat\nabla^0_\alpha q^+_b\right)^{\underline{B}}_{A} \nn \\
&&-i f^{--(ab)}\left[(W^{++}_L)_A^D\, (q^{+}_b)^{\underline{B}}_{D}
- (W^{++}_R)^{\underline{B}}_{\underline{C}}\, (q^{+}_b)^{\underline{C}}_{A}\right],\lb{lie2}
\eea
where
\be
\omega^\alpha_{(ab)} = 2 \lambda_{(ij)(ab)}\epsilon^{(ij)\alpha}
\ee
and
\be
f^{--(ab)} = 2\lambda^{--(ab)} (\epsilon^{--\alpha}\theta^{++}_\alpha)
- 4\lambda^{0(ab)} (\epsilon^{--\alpha}\theta^0_\alpha)\,.
\ee
The bracket for $(V^{++}_R)^{\underline{A}}_{\underline{B}}$ has a form quite analogous to \p{lie1}.

First terms in \p{lie1}, \p{lie2} are just the transformations of the hidden ${\cal N}=3$ supersymmetry which extends
the manifest one to ${\cal N}=6\,$. The remaining terms are reduced on shell to a field-dependent gauge transformation.
Indeed, with making use of the equations of motion \p{Qeq}, \p{VEqs}
these ``superfluous'' terms in the bracket transformations of $V^{++}_L$, $V^{++}_R$ and $q^{+a}$ can be represented, respectively, as
\be
-(\nabla^{++}\Lambda)_A^B\,, \quad -(\nabla^{++}\tilde\Lambda)^{\underline{A}}_{\underline{B}}\,, \quad
\Lambda_A^D\,(q^{+ a})^{\underline{B}}_{D} - \tilde{\Lambda}^{\underline{B}}_{\underline{D}}(q^{+ a})^{\underline{D}}_{A}\,,
\ee
where
\be
\Lambda_A^B := -\frac{4\pi}{k}\,f^{--}_{(ab)}\,(q^{+ a})^{\underline{C}}_{A}\,(\bar q^{+ b})^{B}_{\underline{C}}\,, \quad
\tilde{\Lambda}^{\underline{B}}_{\underline{A}} := -\frac{4\pi}{k}\,f^{--}_{(ab)}\,(q^{+ a})^{\underline{B}}_{C}\,(\bar q^{+ b})^{C}_{\underline{A}}\,.
\ee

Thus, the transformations of hidden supersymmetries can be equivalently derived  as an essential part of the Lie bracket
of the explicit $\cN{=}3$ supersymmetry with the hidden internal automorphisms transformations (i.e.\
the part which retains on shell and
is not reduced to a gauge transformation).

\subsection{Scalar potential}
\label{scal-poten}
One of the basic features of the ABJM model is the sextic potential of the
scalar fields. In \cite{ABJM} it was presented
in the manifestly $\textrm{SU}(4)$ invariant form. Following the ABJM
terminology, there are four complex scalars, two of which, $A_1$ and
$A_2\,$, are in the bifundamental representation while the other two,
$B_1$ and $B_2\,$, are in the anti-bifundamental representation. These scalars
are combined into the $\textrm{SU}(4)$ spinors (``quark'' and ``anti-quark''):
\be
C_I=(A_1,A_2,B_1^\dag,B_2^\dag),\qquad
C^{\dag I}=(A_1^\dag,A_2^\dag,B_1,B_2).
\label{C-ABJM}
\ee
In terms of these quantities the scalar potential is written as
\bea
V_{(ABJM)}&= &\frac{4\pi^2}{k^2}\tr\left(
\frac13\tr(C_I C^{\dag I}C_J C^{\dag J}C_{K}C^{\dag K}
+\frac13 C_I C^{\dag J}C_{J}C^{\dag K}C_K C^{\dag I}\right. \nn\\
&& \left. -2C_IC^{\dag I}C_J C^{\dag K}C_K C^{\dag J}
+\frac43C_I C^{\dag K}C_JC^{\dag I}C_K C^{\dag J}\right).
\label{V-ABJM}
\eea

In our $\cN{=}3$ harmonic superspace formulation of the ABJM model the action
(\ref{N6}) contains no explicit superfield potential, it involves only minimal gauge
interactions of the hypermultiplets with the gauge superfields. As was already mentioned, such a form
of the action is uniquely prescribed by $\cN{=}3$ superconformal invariance.

Here we demonstrate that the scalar potential (\ref{V-ABJM})
naturally stems from the superfield action (\ref{N6}) as a result of
elimination of auxiliary fields.

Upon reducing the action (\ref{N6}) to the component form, the
contributions to the scalar potential come from both the
hypermultiplet and Chern-Simons actions (\ref{Sgauge}), (\ref{hyp3}).
To derive the scalar potential, we take $V^{++}_{L, R}$ in the Wess-Zumino gauge
\p{V++} and discard there gauge fields and all fermionic fields. Further, based on the dimensionality reasoning,
we single out those auxiliary bosonic fields in the hypermultiplet superfields which
are relevant to forming the on-shell scalar potential. As a result we find that it suffices to leave only
the following component fields:
\bea
V^{++}_{L,R}&=&3(\theta^{++})^2u^-_k u^-_l \phi^{kl}_{L,R}
+3i(\theta^{++})^2(\theta^0)^2u^-_ku^-_l X^{kl}_{L,R}+\ldots,\nn\\
q^{+a}&=&u^+_if^{i\,a}+(\theta^0)^2 g^{i\,a}u^+_i
+(\theta^{++}\theta^0)h^{i\,a}u^-_i+\ldots,\nn\\
\bar q^{+}_a&=&-u^+_i\bar f^{i}_a+(\theta^0)^2\bar g^{i}_au^+_i
+(\theta^{++}\theta^0)\bar h^{i}_au^-_i+\ldots.
\label{comps}
\eea
Now, using the component structure of the Chern-Simons action
(\ref{CS-comp}) we can explicitly write down those terms in the action
(\ref{Sgauge}) which are responsible for the scalar potential,
\be
S_{gauge}=-\frac{ik}{6\pi}\tr\int
d^3x(\phi^m_{Lk}[\phi^n_{Lm},\phi^k_{Ln}]
-\phi^m_{Rk}[\phi^n_{Rm},\phi^k_{Rn}])
+\frac k{4\pi}
\tr\int d^3x(\phi^{ij}_LX_{Lij}-\phi^{ij}_RX_{Rij})
+\ldots.
\label{pot}
\ee

To find the component structure of the appropriate part of the hypermultiplet action
we eliminate the auxiliary fields $g^{i\,a}$, $h^{i\,a}$
from the equation of motion $\nabla^{++}q^{+a}=0$,
\be
h^{i\,a}=-2g^{i\,a}=2\phi_L^{ij}f^a_{j}
-2f^a_j\phi_R^{ij}
\ee
and substitute the
component expansions (\ref{comps}) into the hypermultiplet action
(\ref{hyp3}). After integration over the Grassmann and harmonic
variables we obtain
\bea
S_{hyp}&=&
\tr\int d^3x[-\bar f_{i\,a}\phi_L^{ij}\phi_{Ljk}f^{k\,a}
+\bar f_{i\,a}\phi_L^{ij}f^{k\, a}\phi_{Rjk}
+\bar f_{i\,a}\phi_{Ljk}f^{k\,a}\phi_R^{ij}
-\bar f_{i\,a} f^{k\,a}\phi_{Rjk}\phi_R^{ij}
\nn\\&&
-i\bar f_{i\,a}X^{ij}_L f^a_j+i\bar f_{i\,a} f^a_{j}X^{ij}_R+\ldots].
\label{222}
\eea
The auxiliary fields $X_{L,R}$ appear in the action $S_{gauge}+S_{hyp}$ as Lagrange multipliers
for the relations
\be
\phi^{ij}_L=\frac{2\pi i}k(f^{i\,a}\bar f^{j}_{a}
 +f^{j\,a}\bar f^{i}_{a}),\qquad
\phi^{ij}_R=-\frac{2\pi i}k(\bar f^{i\,a} f^{j}_{a}
+\bar f^{j\,a} f^{i}_{a}).
\label{phi}
\ee
As the final step, we substitute these expressions for the auxiliary fields back into the
actions (\ref{pot}), (\ref{222}) and, after some simple algebra, obtain
the scalar potential in the following form
\bea
V_{scalar}&=&-\frac{8\pi^2}{3k^2}\tr\{
f^{i\,a}\bar f_{k\,a}
 (f^{j\,b}\bar f_{i\,b}+f^{b}_i\bar f^j_{b})
 (f^{k\,c}\bar f_{j\,c}+f^{c}_j\bar f^k_{c})\nn\\&&
+\bar f^{i\,a} f_{k\,a}
 (\bar f^{j\,b} f_{i\,b}+\bar f^{b}_i f^j_{b})
 (\bar f^{k\,c} f_{j\,c}+\bar f^{c}_j f^k_{c})
\}\nn\\
&&-\frac{4\pi^2}{3k^2}\tr\{
f^{i\,a}\bar f_{k\,a}
 (f^{k\,c}\bar f_{j\,c}+f^{c}_j\bar f^k_{c})
 (f^{j\,b}\bar f_{i\,b}+f^{b}_i\bar f^j_{b})\nn\\&&
+\bar f^{i\,a} f_{k\,a}
 (\bar f^{k\,c} f_{j\,c}+\bar f^{c}_j f^k_{c})
 (\bar f^{j\,b} f_{i\,b}+\bar f^{b}_i f^j_{b})
\}\nn\\
&&
-\frac{4\pi^2}{k^2}\tr\{\bar f_i^{a}(f^{i\,c}\bar f^j_{c}+f^{j\,c}\bar f_{c}^i)
 f^k_{a}(\bar f_j^{b}f_{k\,b}+\bar f^{b}_kf_{j\,b})
\nn\\&&
+\bar f_i^{a}(f^{c}_j\bar f_{k\,c}+f^{c}_k\bar f_{j\,c})
 f^k_{a}(\bar f^{i\,b}f^j_{b}+\bar f^{j\,b}f_{b}^i)\}.
\label{V}
\eea

The potential (\ref{V}) looks rather complicated and its identity
with (\ref{V-ABJM}) is not immediately obvious. To show the coincidence of these two
expressions, we pass to the ABJM notations (\ref{C-ABJM}) by identifying
\bea
f^{i\,a}&=&(f^{11},f^{12},f^{21},f^{22})=
(A_1,A_2,B_1^\dag,B_2^\dag)=C_I,\nn\\
\bar f_{i\,a}&=&(\bar f_{11},\bar f_{12},\bar f_{21},\bar f_{22})
=(A_1^\dag,A_2^\dag,B_1,B_2)=C^{\dag I}.
\label{f-C}
\eea
Substituting (\ref{f-C}) into (\ref{V}) and making appropriate Fierz rearrangements,
after some tedious computation we end up with the desired identity
\be
V_{scalar}=V_{(ABJM)}.
\ee

Thus we have explicitly shown that the scalar potential derived from the
superfield action (\ref{N6}) is just the potential found by ABJM \cite{ABJM}.
We point out once more that in our $\cN{=}3$ superfield
formulation the scalar potential emerges
solely as a result of elimination of auxiliary fields, without any presupposed superfield potential.
The other interaction terms in the ABJM
model (e.g., the quartic interaction of two scalars with two
fermions, etc) originate from \p{N6} in a similar way.

\setcounter{equation}{0}

\section{Other options}
Here we discuss some other choices of the gauge group and/or of the representation
of the hypermultiplet superfields admitting additional hidden supersymmetries and R-symmetries.

\subsection{The $(N,M)$ model}

The $\cN{=}6$ supersymmetry and $\textrm{SO}(6)$ R-symmetry in the $(N,M)$ model corresponding to the choice \p{bi2}
can be proved following the same line as in the case of $(N, \bar{M})$ model.
The hypermultiplet action is
\bea
&&S^\prime_{hyp}= \int d\zeta^{(-4)}(\bar{q}^+_a)^{A\underline{A}}\nabla^{++}
(q^{+a})_{A\underline{A}}\,,
\eea
where
\bea
&&(\nabla^{++}q^{+a})_{A\underline{A}}=\cD^{++}(q^{+a})_{A\underline{A}}
+ (V^{++}_L)_A^B (q^{+a})_{B\underline{A}}+(V^{++}_R)_{\underline{A}}^{\underline{B}}
(q^{+a})_{A\underline{B}}\,.
\eea
The additional three supersymmetry transformations are
\bea
\delta_\epsilon (q^{+a})_{A\underline{A}}=i\epsilon^{(ab)\alpha}\hat{\nabla}{}^0_\alpha
(q^+_b)_{A\underline{A}}\,, \lb{qext2}
\eea
\bea
\delta_\epsilon (V^{++}_L)_A^B= \frac{8\pi}k\epsilon^{\alpha\,(ab)}
 \theta^0_\alpha (q^+_a)_{A\underline{B}}(\bar q^+_b)^{B\underline{B}}\,,\quad
\delta_\epsilon (V^{++}_R)_{\underline{A}}^{\underline{B}}=
-\frac{8\pi}k\epsilon^\alpha_{(ab)}
 \theta^0_\alpha (q^{+a})_{B\underline{A}}(\bar q^{+b})^{B\underline{B}}\,, \lb{Vext2}
\eea
where now
\bea
\hat{\nabla}{}^0_\alpha (q^+_b)_{A\underline{A}} &=& D^0_\alpha (q^+_b)_{A\underline{A}}
+[(V^0_{L\alpha})_A^B+\theta^{--}_\alpha (W^{++}_L)_A^B](q^+_b)_{B\underline{A}} \nn \\
&& +\, [(V^0_{R\alpha})_{\underline{A}}^{\underline{B}}
+\theta^{--}_\alpha (W^{++}_R)_{\underline{A}}^{\underline{B}}](q^+_b)_{A\underline{B}}\}\,.
\eea
The transformations of the hidden $\textrm{SO}(6)/[\textrm{SU}(2)_R\times \textrm{SU}(2)_{ext}]$ R-symmetry mimic eqs.\ \p{non-ab}--\p{var12},
the only difference consists in that the variation $\delta_\lambda V^{++}_R$ has the opposite sign as compared
to $\delta_\lambda V^{++}_R$, like in \p{Vext2}. The invariance of the total gauge-hypermultiplet action
is checked as in the previously considered $(N, \bar M)$ model. The cancelation of the quartic terms in the full
variations of the hypermultiplet action is a consequence of the identity similar to \p{BasId}.

\subsection{$\textrm{SU}(N)\times \textrm{SU}(N)$ model}

Let us come back to the hypermultiplet superfield $(N,\bar{M})$ model
(\ref{bi1}), choose there $N=M$ and restrict the gauge group to be $\textrm{SU}(N)\times \textrm{SU}(N)$.
The hidden $\cN{=}6$ supersymmetry transformations \p{epsilon4},
as well as the $\textrm{SO}(6)$ transformations \p{var12}, should be slightly modified in this case
in order to obey the tracelessness restrictions of the gauge group.
In particular, eqs.\ \p{epsilon4} change as
\bea
&& \delta_\epsilon V^{++}_L= \frac{8\pi}k\epsilon^{\alpha(ab)}
 \theta^0_\alpha \left(q^+_a\bar q^+_b-\frac{1}{N}\tr q^+_a\bar q^+_b\right), \nn \\
&& \delta_\epsilon V^{++}_R= \frac{8\pi}k\epsilon^{\alpha(ab)}
 \theta^0_\alpha \left(\bar q^+_a q^+_b-\frac{1}{N}\tr q^+_a\bar q^+_b\right)
\eea
(equations of motion for $V^{++}_L, V^{++}_R$ \p{VEqs} undergo a similar modification).
Analogous tracelessness conditions should be imposed on the hidden $\cN{=}6$ supersymmetry and $\textrm{SO}(6)$ R-symmetry
transformations of the $(N,M)$ model \p{bi2} restricted to $N=M$ and to the gauge group $\textrm{SU}(N)\times \textrm{SU}(N)$.

In both cases the total gauge - hypermultiplet actions remain invariant because the unwanted quartic terms in
the variations of the hypermultiplet actions induced by $\delta V^{++}_L$ and $\delta V^{++}_R$ vanish as
in the original $\tU(N)\times \tU(M)$ settings.
Note that the restriction of two U(1) factors  in $\tU(N)\times\tU(M)$
to the diagonal $\tU(1)$ does not break the hidden ${\cal N}{=}6$ supersymmetry
and $\textrm{SO}(6)$ R-symmetry for both the $(N,M)$ and $(N,\bar{M})$
models since the trace parts in the variations (\ref{epsilon4}),
(\ref{var12}) and (\ref{Vext2})
coincide and cancel each other in the appropriate variations of the
hypermultiplet action. As a result, the gauge group
$\textrm{SU}(N)\times\textrm{SU}(M)\times\tU(1)$
is the admissible option for the existence of ${\cal N}{=}6$
supersymmetry and $\textrm{SO}(6)$ R-symmetry, in agreement with the conclusion made in
\cite{STac}.
 On the contrary, when restricting the gauge group
to $\textrm{SU}(N)\times \textrm{SU}(M)$, $N\neq M$, there survive quartic terms $\sim (1/N - 1/M)$ in these variations and there
is no way to cancel them. Thus for the gauge group $\textrm{SU}(N)\times \textrm{SU}(M)$ both models have neither
hidden supersymmetry nor hidden R-symmetry.

The $(N,\bar{N})$ model for the gauge group $\textrm{SU}(N)\times \textrm{SU}(N)$
is invariant under the P-parity transformations
\bea
&&P V^{++}_L(\zeta_P)=V^{++}_R(\zeta),\quad P W^{++}_L(\zeta_P)=-W^{++}_R,\nn\\
&&P (q^{+a})_A^{\underline{B}}(\zeta_P)=(\bar{q}^{+a})_{\underline{A}}^B(\zeta).
\eea
The $(N, N)$ model for the group $\textrm{SU}(N)\times \textrm{SU}(N)$ also respects P-parity
which, on the hypermultiplets,  is represented by the following
transformations
\bea
P (q^{+a})_{A\underline{B}}(\zeta_P)=
(q^{+a})_{B\underline{A}}(\zeta)\,,\quad P (\bar{q}^+_a)^{A\underline{B}}(\zeta_P)=(\bar{q}^+_a)^{B\underline{A}}(\zeta)\,.
\eea

\subsection{$\textrm{O}(N)\times \textrm{USp}(2M)$ model}
At the component and $\cN{=}2$ superfield level, this interesting option was proposed in \cite{Hosomichi,ABJ,Li}. Here we treat
it within the $\cN{=}3$ harmonic superfield formalism.

Let us consider the hypermultiplet matrix superfield
\bea
(q^{+a})_A^{\underline{A}},
\eea
where $A=1,\ldots, N$ is the vector index of the real $\textrm{SO}(N)$ group and
$\underline{A}=1,\ldots, 2M$ is the spinor index of the $\textrm{USp}(2M)$
group which is defined as a subgroup in $\tU(2M)$ such that it  preserves
the skew-symmetric metric
\be
\Omega_{\underline{A}\underline{B}},\quad \Omega^{\underline{A}\underline{B}} = -\overline{(\Omega_{\underline{A}\underline{B}})}\,, \quad
\Omega_{\underline{A}\underline{B}}
\Omega^{\underline{B}\underline{C}}=\delta_{\underline{A}}^{\underline{C}}\,. \lb{ConjOmega}
\ee
This metric, like $\varepsilon_{ik}$ in the $\textrm{SU}(2) = \textrm{USp}(2)$ case, can be used to raise or lower
the fundamental representation indices $\underline{A}$. The index $a=1,2$ is treated as the global $\textrm{SO}(2)$ one in this case.
The corresponding $\widetilde{\;\;}$ conjugation rules for hypermultiplets are
\be
\widetilde{[(q^{+a})_A^{\underline{A}}]}= \Omega_{\underline{A}\underline{B}}\,({q}^{+a})_{A}^{\underline{B}}\,.
\ee
Taking into account the definitions \p{ConjOmega}, this pseudoreality condition is compatible with the property that
the $\widetilde{\;\;}$ conjugation for the $q^+$ superfields squares to $-1\,$\footnote{Giving up the pseudoreality
condition amounts to a non-minimal variant with two copies of the pseudoreal $q^{+ a}$. This enlargement
of the field representation ($16NM$ real physical scalar as compared to $8NM$ in the case with the pseudoreality
condition) does not introduce any new feature.}. The gauge-covariantized harmonic derivative is defined as
\bea
&& \nabla^{++}(q^{+a})_A^{\underline{A}} ={\cal D}^{++}(q^{+a})_A^{\underline{A}} + (V_L^{++})_{AB}(q^{+a})_B^{\underline{A}}
- (V^{++}_R)^{\underline{A}}_{\underline{B}}(q^{+a})_A^{\underline{B}}\,, \nn \\
&& (V_L^{++})_{AB} = -(V_L^{++})_{BA}\,, \quad
\Omega_{\underline{D}\underline{A}}\,(V^{++}_R)^{\underline{A}}_{\underline{B}} =
\Omega_{\underline{B}\underline{A}}\,(V^{++}_R)^{\underline{A}}_{\underline{D}}\,.
\eea

This gauge group assignment of the hypermultiplet superfields  is compatible
with only two additional supersymmetry  transformations
\bea
\delta_\epsilon (q^{+a})_A^{\underline{A}}=\epsilon^{\alpha\,(ab)}
\hat\nabla^0_\alpha (q^{+b})_A^{\underline{A}}\,,\lb{qext3}
\eea
where
\be
\epsilon^{\alpha\,(ab)}=\epsilon^\alpha_1(\tau_1)^{ab}+
\epsilon^\alpha_3(\tau_3)^{ab}\,,
\ee
$\epsilon^\alpha_1$ and $\epsilon^\alpha_3$ being real spinors
(i.e.\ $\epsilon^{\alpha\,(ab)}$ is the rank 2 symmetric traceless
$\textrm{SO}(2)$ tensor), $\tau_1$ and $\tau_3$ are Pauli matrices.
The two additional transformations of the $\textrm{SO}(N)\times \textrm{USp}(2M)$
prepotentials have the form
\be
\delta_\epsilon (V^{++}_L)_{AB}= -\frac{8i\pi}{k}\epsilon^{\alpha (ab)}
(q^{+a})_A^{\underline{B}}({q}^{+b})_{B\underline{B}},\quad
\delta_\epsilon (V^{++}_R)_{\underline{A}}^{\underline{B}}
=-\frac{8i\pi}{k}\epsilon^{\alpha (ab)}(q^{+a})_B^{\underline{B}}
({q}^{+b})_{B\underline{A}}. \lb{Vext3}
\ee

The total $\textrm{SO}(N)\times \textrm{USp}(2M)$ Chern-Simons-hypermultiplet action
\be
S = S_{CS}(V_L^{++}) - S_{CS}(V^{++}_R) + \int d\zeta^{(-4)}\, q^{+a}_{A\underline{A}}\nabla^{++}q^{+a\underline{A}}_A \lb{ActSp}
\ee
is invariant under full $\cN{=}5$ supersymmetry involving the manifest off-shell $\cN{=}3$ supersymmetry  and
the above two additional on-shell ones.

The action \p{ActSp} is also invariant under the following hidden R-symmetry transformation
\bea
\delta_\lambda q^{+a\underline{A}}_A = [ \lambda^{0(ab)}- \lambda^{++(ab)}\hat\nabla^{--}
- 2\lambda^{--(ab)}\theta^{++ \alpha}\hat\nabla^0_\alpha
+ 4\lambda^{0(ab)}\theta^{0 \alpha}\hat\nabla^0_\alpha]q^{+b\underline{A}}_A\,, \lb{RSp1} \\
\delta_\lambda (V^{++}_L)_{AB}= \frac{4i\pi}{k}\varphi^{(ab)}
(q^{+a})_A^{\underline{B}}({q}^{+b})_{B\underline{B}},\;
\delta_\lambda (V^{++}_R)_{\underline{A}}^{\underline{B}}
= \frac{4i\pi}{k}\varphi^{(ab)}(q^{+a})_B^{\underline{B}}
({q}^{+b})_{B\underline{A}}\,, \lb{RSp}
\eea
where
\be
\varphi^{(ab)} = 4\lambda^{--(ab)}(\theta^{++\alpha}\theta^0_\alpha) - 8 \lambda^{0(ab)}(\theta^0)^2
\ee
and the $\textrm{SO}(2)$ index $(ab)$ refers to the rank 2 symmetric traceless $\textrm{SO}(2)$ tensor. These transformations, modulo equations
of motion for auxiliary fields and field-dependent gauge transformations, together with those of the groups $\textrm{SU}(2)_c$ and $\textrm{SO}(2)$,
form the 10-parameter $\textrm{SO}(5)$ R-symmetry (3 parameters of $\textrm{SU}(2)_c$ plus 1 parameter of $\textrm{SO}(2)$ plus 6 real parameters
$\lambda^{(ik)(ab)}$ of \p{RSp1}, \p{RSp}). The commutator of \p{RSp} with the explicit $\cN{=}3$ supersymmetry \p{Qq} yields (once again,
on-shell and up to a field-dependent gauge transformation) just the hidden $\cN{=}3$ supersymmetry \p{qext3}, \p{Vext3}. The
reason why the parameters $\epsilon^{\alpha(ab)}$ and $\lambda^{(ik)(ab)}$ should be symmetric in the $\textrm{SO}(2)$ indices
$a, b$ is the requirement that the variation of the hypermultiplet part of the action \p{ActSp} with respect to \p{RSp1}
can be compensated, modulo a total derivative, by the appropriate variation of the Chern-Simons actions. The further restriction
that these parameters are traceless in $a, b$ arises as the condition of vanishing of the unwanted quartic terms in the full group
variation of the hypermultiplet action. After the appropriate Fierz rearrangement, these terms (with the infinitesimal transformation
parameters $\epsilon^{\alpha(ab)}$ or $\lambda^{(ik)(ab)}$ detached) are gathered into the structure
\be
(\epsilon^{cg}\,q^{+ c}_{B\underline{A}}q^{+ g}_{A\underline{D}})\,(\epsilon^{d(a}\,q^{+ b)}_{B}{}^{\underline{D}} q^{+ d}_{A}{}^{\underline{A}})\,.
\ee
It is easy to check that the $(ab)$ trace part of this expression is not vanishing and cannot be canceled with any modification
of the (super)group transformations, while the traceless part is identically zero. So the
transformation parameters should be symmetric traceless.

There exist some other choices of the gauge groups and/or the representation assignments of the hypermultiplet matter which seemingly admit
extra supersymmetries and R-symmetries (see e.g. \cite{Li}). We are planning to discuss them elsewhere. The cancellation of the
quartic terms in the variation of the $\cN{=}3$ superfield hypermultiplet action seems to be a simple powerful criterion
for selecting all non-trivial possibilities.

The component forms of the Chern-Simons - hypermultiplet superfield actions considered in this Section, in particular, the corresponding
sextic scalar potentials, can be derived in the same way as for the $\tU(N)\times \tU(M)$ model in Subsection 3.4.
\setcounter{equation}{0}

\section{Models with $\cN{=}8$ supersymmetry}
As claimed in \cite{ABJM}, in the case of $\textrm{SU}(2)\times \textrm{SU}(2)$ gauge group the ABJM model has an enhanced $\cN{=}8$ supersymmetry
and coincides with the $\textrm{SO}(4)$ BLG model \cite{BLG,Gustavsson}.
Here we show this using the $\cN{=}3$ superfield formalism.

We start from the particular $\textrm{SU}(2)\times \textrm{SU}(2)$ case of the general $\tU(N)\times \tU(M)$ action \p{N6}:
\bea
S_{su(2)}= S_{CS}[V^{++}_L]-S_{CS}[V^{++}_R] -\int d\zeta^{(-4)}\bar q^{+}_{a}{}^{A\un{A}}\nabla^{++} q^{+a}_{A\un{A}}\,, \quad
\widetilde{(q^{+a}_{A\un{A}})} = - \bar{q}_a^+{}^{A\un{A}}\,,
\label{hyp5}
\eea
where we have written down the doublet indices of both gauge $\textrm{SU}(2)$ groups on the same level, using the equivalency of the fundamental
representation of $\textrm{SU}(2)$ and its conjugate. In this notation, the covariant derivative $\nabla^{++} q^{+a}_{A\un{A}}$ is written as
\be
\nabla^{++} q^{+a}_{A\un{A}} = {\cal D}^{++} q^{+a}_{A\un{A}} +(V^{++}_L)_A^B q^{+a}_{B\un{A}} + (V^{++}_R)_{\un{A}}^{\un{B}} q^{+a}_{A\un{B}}\,.
\lb{Nablasu2}
\ee

Now we give up the notation in which $\textrm{SU}(2)_{ext}$ symmetry acting on the index $a$ is manifest and will treat  the superfields
$ q^{+1}_{A\un{B}}$ and  $q^{+2}_{A\un{B}}$ separately. Either these superfields, together with their $\widetilde{\;\;}$ conjugates, can be combined
into two independent pseudo-real doublets of two Pauli-G\"ursey $\textrm{SU}(2)$ groups \cite{book}:
\bea
&& \textrm{SU}(2)_{PG\,I}: \quad q^{+ \hat{a}}_{A\un{B}} := ( q^{+1}_{A\un{B}}, \; \bar q^{+}_{1\,A\un{B}})\,, \qquad
\widetilde{(q^{+ \hat{a}}_{A\un{B}})}= - q^{+}_{\hat{a}}{}^{A\un{B}}\,, \nn \\
&& \textrm{SU}(2)_{PG\,II}: \quad q^{+ \check{a}}_{A\un{B}} := ( q^{+2}_{A\un{B}}, \; \bar q^{+}_{2\,A\un{B}})\,, \qquad
\widetilde{(q^{+ \check{a}}_{A\un{B}})}= - q^{+}_{\check{a}}{}^{A\un{B}}\,. \lb{PGgr}
\eea
In this new notation the action \p{hyp5} is rewritten as
\bea
&& S_{su(2)} = S_{CS}[V^{++}_L]-S_{CS}[V^{++}_R] - \frac{1}{2} \int d\zeta^{(-4)}\left({\cal L}_I + {\cal L}_{II}\right), \lb{ActPG} \\
&& {\cal L}_I =q^{+}_{\hat{a}}{}^{A\un{B}}\nabla^{++} q^{+\hat{a}}_{A\un{B}}\,, \qquad {\cal L}_{II} =
q^{+}_{\check{a}}{}^{A\un{B}}\nabla^{++} q^{+\check{a}}_{A\un{B}}\,. \lb{LgPG}
\eea
The covariant derivative $\nabla^{++}$ acts in the same way as in \p{Nablasu2}.

The rearranged action \p{ActPG} manifests three mutually commuting off-shell $\textrm{SU}(2)$ symmetry: two Pauli-G\"ursey
symmetries $\textrm{SU}(2)_{PGI}$ and $\textrm{SU}(2)_{PGII}$
realized on the hypermultiplet doublet indices $\hat{a}$ and $\check{a}$, as well as the standard automorphism $\textrm{SU}(2)_R$ symmetry (or $\textrm{SU}(2)_c$
symmetry which is indistinguishable from $\textrm{SU}(2)_R$ on physical fields). The original $\textrm{SU}(2)_{ext}$ symmetry is of course also there, but
in the new formulation it is realized in some implicit way. The gauge group $\textrm{SU}(2)\times \textrm{SU}(2)$ commutes with all these symmetries. The possibility
to pass to two independent pseudo-real hypermultiplet superfields is directly related to the fact that the original complex hypermultiplet superfields
and their conjugates prove to be in the same bifundamental representation of $\textrm{SU}(2)\times \textrm{SU}(2)$ because of the equivalency of the fundamental
representation and its conjugate in the $\textrm{SU}(2)$ case. Just due to this property one can combine them into the $\textrm{SU}(2)_{PG}$ doublets as in \p{PGgr}.
As we shall see soon, this possibility is crucial for the existence of the hidden $\cN{=}8$ supersymmetry and $\textrm{SO}(8)$ R-symmetry
in the $\textrm{SU}(2)\times \textrm{SU}(2)$ model. In the case of the gauge supergroup $\textrm{SU}(N)\times \textrm{SU}(N), N \geq 3\,,$ the hypermultiplets and
their conjugates transform according to non-equivalent representations and therefore cannot be joined into $\textrm{SU}(2)_{PG}$ doublets
(neither for the $(N, \bar{N})$ model nor for the $(N, N)$ one). Correspondingly, their $\cN{=}6$ supersymmetry and $\textrm{SO}(6)$ R-symmetry
are not further enhanced. The same is true for $\tU(N)\times \tU(M)$ models including
$\tU(2)\times \tU(2)$ and $\tU(1)\times \tU(1)$ ones. We shall see that there exists an extended version of the
$\tU(1)\times \tU(1)$ model which still admits $\textrm{SU}(2)_{PG}$ doublet structure (it involves 8 complex physical scalar fields  instead of 4 such fields in
the minimal $\tU(1)\times \tU(1)$ case). It is obtained as a reduction of the $\textrm{SU}(2)\times \textrm{SU}(2)$ model and also possesses $\cN =8$ supersymmetry
and $\textrm{SO}(8)$ R-symmetry. The formulation in terms of two pseudo-real hypermultiplets exists as well in the more general case of gauge groups
$\textrm{USp}(2N)\times \textrm{USp}(2M)$ for which the bifundamental representation $(2N, 2M)$ is also equivalent to its complex conjugate due to the existence
of the invariant skew-symmetric metrics. For generic values of $N$ and $M$, however, no hidden supersymmetries
or full $\textrm{SO}(8)$
R-symmetry arise in the $\textrm{USp}(2N)\times \textrm{USp}(2M)$ models as we argue below.

In revealing the hidden symmetries inherent in the action \p{ActPG} we start with the R-symmetries. The most evident extra symmetry
is realized by linear transformations
\be
\delta_\lambda q^{+\hat{a}}_{A\un{A}} = \lambda^{\hat{a} \check{a}}q^{+}_{\check{a}}{}_{A\un{A}}\,, \quad
\delta_\lambda  q^{+\check{a}}_{A\un{A}} = \lambda^{\hat{a} \check{a}}q^{+}_{\hat{a}}{}_{A\un{A}}\,, \quad
\delta_\lambda V^{++}_L = \delta_\lambda V^{++}_R = 0\,,\lb{Simple}
\ee
where $\lambda^{\hat{a} \check{a}}$ are four real parameters. They commute with the manifest $\cN{=}3$ supersymmetry and close off shell
on the product $\textrm{SU}(2)_{PGI}\times \textrm{SU}(2)_{PGII} = \textrm{SO}(4)_{PG}$. Together with the latter they generate $\textrm{SO}(5)$ symmetry which is the
maximal subsymmetry of the full R-symmetry group of the model under consideration which commutes with the manifest $\cN{=}3$ supersymmetry.

Two other sets of the hidden internal symmetries are represented by the transformations of the form which we already met
in the examples considered earlier.

The first set of additional transformations leaving the action \p{ActPG} invariant is as follows
\bea
&& \delta_\omega q^{+\hat{a}}_{A\un{A}} =[\omega^{0}- \omega^{++}\hat\nabla^{--}
- 2\omega^{--}\theta^{++ \alpha}\hat\nabla^0_\alpha
+ 4\omega^{0}\theta^{0 \alpha}\hat\nabla^0_\alpha]q^{+\hat{a}}_{A\un{A}}\,, \nn \\
&& \delta_\omega q^{+\check{a}}_{A\un{A}} = -[\omega^{0}- \omega^{++}\hat\nabla^{--}
- 2\omega^{--}\theta^{++ \alpha}\hat\nabla^0_\alpha
+ 4\omega^{0}\theta^{0 \alpha}\hat\nabla^0_\alpha]q^{+\check{a}}_{A\un{A}}\,, \nn \\
&& \delta_\omega (V^{++}_L)^A_B = \frac{2i\pi}{k}\varphi \left(q^{+\check{a}}_{B\un{A}}q^{+A\un{A}}_{\check{a}}
-q^{+\hat{a}}_{B\un{A}}q^{+A\un{A}}_{\hat{a}}\right), \nn \\
&& \delta_\omega (V^{++}_R)^{\un{A}}_{\un{B}} = \frac{2i\pi}{k}\varphi \left(q^{+\hat{a}}_{A\un{B}}q^{+A\un{A}}_{\hat{a}}
-q^{+\check{a}}_{A\un{B}}q^{+A\un{A}}_{\check{a}}\right), \lb{1set}
\eea
where
\be
\varphi = 4\omega^{--}(\theta^{++\alpha}\theta^0_\alpha) - 8 \omega^0(\theta^0)^2
\ee
and $\omega^0 = \omega^{(ik)}u^+_iu^-_k, \omega^{\pm\pm} = \omega^{(ik)}u^{\pm}_iu^{\pm}_k\,$. The cancelation
of the quartic terms in the variation of the hypermultiplet action comes about under the two conditions of the same type
\be
(q^{+\hat{b}}_{A\un{B}}\,q^{+}_{\hat{b}\,B\un{A}})\,q_{\hat{a}}^{+ B\un{B}} = 0\,, \quad
(q^{+\check{b}}_{A\un{B}}\,q^{+}_{\check{b}\,B\un{A}})\,q_{\check{a}}^{+ B\un{B}} = 0\,, \lb{CondSU2}
\ee
which are easily checked to be satisfied for the $\textrm{SU}(2)$ case. These transformations in their on-shell closure yield
the conformal R-symmetry group $\textrm{SU}(2)_c$. Taken together with the $\textrm{SU}(2)_c$ transformations, they amount to two independent
$\textrm{SU}(2)$ rotations of the physical fields in $q^{+\hat{b}} = f^{i \hat{b}}u^+_i + \ldots $ and
$q^{+\check{b}} = f^{i \check{b}}u^+_i + \ldots $ with respect to their harmonic indices $i$.

The last set of hidden R-symmetry transformations is given by
\bea
&& \delta_\sigma q^{+\hat{a}}_{A\un{A}} =[\sigma^{0 \,\hat{a}\check{b}}- \sigma^{++\,\hat{a}\check{b}}\hat\nabla^{--}
- 2\sigma^{--\,\hat{a}\check{b}}\theta^{++ \alpha}\hat\nabla^0_\alpha
+ 4\sigma^{0\,\hat{a}\check{b}}\theta^{0 \alpha}\hat\nabla^0_\alpha]\,q^{+}_{\check{b}\,A\un{A}}\,, \nn \\
&& \delta_\sigma q^{+\check{a}}_{A\un{A}} = -[\sigma^{0\,\hat{a}\check{b}}- \sigma^{++\,\hat{a}\check{b}}\hat\nabla^{--}
- 2\sigma^{--\,\hat{a}\check{b}}\theta^{++ \alpha}\hat\nabla^0_\alpha
+ 4\sigma^{0\,\hat{a}\check{b}}\theta^{0 \alpha}\hat\nabla^0_\alpha]\,q^{+}_{\hat{b}\,A\un{A}}\,, \nn \\
&& \delta_\sigma (V^{++}_L)_{AB} = -\frac{4i\pi}{k}\varphi^{\hat{a}\check{b}}\,q_{\hat{a} (A}^{+ \un{B}}\,q^{+}_{\check{b}\,B)\un{B}}\,, \;
\delta_\sigma (V^{++}_R)_{\un{A}\un{B}} = \frac{4i\pi}{k}\varphi^{\hat{a}\check{b}}\, q_{\hat{a} (\un{A}}^{+ B}\, q^{+}_{\check{b} B \un{B})}\,,
\lb{2set}
\eea
with
\be
\varphi^{\hat{a}\check{b}} = 4\sigma^{--\,\hat{a}\check{b}}(\theta^{++\alpha}\theta^0_\alpha) - 8 \sigma^{0\,\hat{a}\check{b}}(\theta^0)^2
\ee
and $\sigma^{++\hat{a}\check{b}} = \sigma^{(ik)\hat{a}\check{b}}u^+_iu^+_k$, etc. The conditions of vanishing of the relevant quartic terms in the
variation of hypermultiplet action are again \p{CondSU2}.

The total number of parameters of all R-symmetries of the action \p{ActPG} is a sum of 12 parameters of four commuting $\textrm{SU}(2)$ symmetries
including \p{1set}, of 4 parameters
of the transformations \p{Simple} and of 12 parameters of the transformations \p{2set}, i.e.\ total of 28 parameters,
the dimension of the group $\textrm{SO}(8)$. Indeed, one can check that all these R-symmetry transformations close modulo field-dependent gauge
transformations and superfield equations of motion, and their closure is just the maximal R-symmetry group $\textrm{SO}(8)$.

Commuting \p{1set} and \p{2set} with the transformations of the manifest $\cN{=}3$ supersymmetry, we find 5 new hidden supersymmetries,
with the Lie bracket parameters $\varepsilon_\alpha \propto \omega^{(ik)}\epsilon_{\alpha (ik)}$ and
$\varepsilon_\alpha^{\hat{a}\check{b}}\propto \sigma^{(ik)\,\hat{a}\check{b}}\epsilon_{\alpha (ik)}$.
They are realized by the following transformations
\bea
&& \delta_{\varepsilon}q^{+\hat{a}}_{A\un{B}} =
\varepsilon^{\alpha}\hat{\nabla}^0_\alpha q^{+\hat{a}}_{A\un{B}}+
\varepsilon^{\alpha \,\hat{a}\check{b}}\hat{\nabla}^0_\alpha q^{+}_{\check{b}\,A\un{B}}\,,
\quad
\delta_{\varepsilon}q^{+\check{a}}_{A\un{B}} =
 -\varepsilon^{\alpha}\hat{\nabla}^0_\alpha q^{+\check{a}}_{A\un{B}}
 -\varepsilon^{\alpha \,\hat{b}\check{a}}\hat{\nabla}^0_\alpha q^{+}_{\hat{b}\,A\un{B}}\,, \nn \\
&&\delta_\varepsilon (V^{++}_L)_{AB} = -\frac{4i\pi}{k}(\varepsilon^{\alpha}\theta^0_\alpha)\left(q^{+\check{a}}_{B\un{A}}q^{+ \un{A}}_{\check{a} A}
-q^{+\hat{a}}_{B\un{A}}q^{+ \un{A}}_{\hat{a}A}\right) +\frac{8i\pi}{k}(\varepsilon^{\alpha\hat{a}\check{b}}\theta^0_\alpha)
\,q_{\hat{a} (A}^{+ \un{B}}\,q^{+}_{\check{b}\,B)\un{B}}\,, \nn \\
&& \delta_\varepsilon (V^{++}_R)_{\un{A}\un{B}} = \frac{4i\pi}{k}(\varepsilon^{\alpha}\theta^0_\alpha)
\left(q^{+\hat{a}}_{A\un{B}}q^{+ A}_{\hat{a}\,\un{A}}-q^{+\check{a}}_{A\un{B}}q^{+ A}_{\check{a}\,\un{A}}\right) -
\frac{8i\pi}{k}(\varepsilon^{\alpha\hat{a}\check{b}}\theta^0_\alpha)\, q_{\hat{a} (\un{A}}^{+ B}\, q^{+}_{\check{b} B \un{B})}\,.\lb{SUSYsu2}
\eea
Together with the manifest $\cN{=}3$ supersymmetry these five extra ones form the $\cN{=}8$ supersymmetry, with an on-shell closure.
Since the action \p{ActPG} is $\cN{=}3$ superconformal, it is also $\cN{=}8$ superconformal.

We close this Section with two comments.

First, the reason why the models with the $\textrm{USp}(2N)\times \textrm{USp}(2M)$ gauge group have neither additional supersymmetries
nor full $\textrm{SO}(8)$ R-symmetry, despite
their formal resemblance to the $\textrm{SU}(2)\times \textrm{SU}(2)$ model, is that the conditions \p{CondSU2} are not satisfied in the generic $N > 1, M >1$ case.
The choice of $N=M=1$ is the unique option when they are valid.
Thus the only additional internal symmetry of the $\textrm{USp}(2N)\times\textrm{USp}(2M)$
models (extending the manifest $\textrm{SO}(4)$ one) is the
$\textrm{SO}(5)/\textrm{SO}(4)$
symmetry (\ref{Simple}) commuting with the ${\cal N}{=}3$ supersymmetry and not
affecting the gauge superfields at all.

Secondly, it is a consistent reduction to put
\be
\mbox{(a)} \;\; q^{+\hat{a}}_{11} = q^{+\hat{a}}_{22} = q^{+\check{a}}_{11} = q^{+\check{a}}_{22} = 0\quad \mbox{or} \quad
\mbox{(b)} \;\; q^{+\hat{a}}_{12} = q^{+\hat{a}}_{21} = q^{+\check{a}}_{12} = q^{+\check{a}}_{21} = 0\;. \lb{Red1}
\ee
These conditions break the gauge group $\textrm{SU}(2)\times \textrm{SU}(2)$ down to its subgroup $\tU(1)\times \tU(1)$.
Actually, two options in \p{Red1} are equivalent to each other and one can focus on (\ref{Red1}a). In this case one is left with
four independent hypermultiplets $q^{+\hat{a}}_{12}, q^{+\hat{a}}_{21}$  and $q^{+\check{a}}_{12}, q^{+\check{a}}_{21}$ as compared
with eight such hypermultiplets in the $\textrm{SU}(2)\times \textrm{SU}(2)$ case and two hypermultiplets in the minimal $\tU(1)\times \tU(1)$ case considered
in Subsection 3.2. The numbers of real scalar fields in these models are, respectively, 16, 32 and 8. The doubling of hypermultiplets as compared to the
minimal $\tU(1)\times \tU(1)$ case allows one to preserve the $\textrm{SU}(2)_{PG}$ multiplet structure and to retain all properties of the
``parent'' $\textrm{SU}(2)\times \textrm{SU}(2)$ model: the $\cN{=}8$ supersymmetry and $\textrm{SO}(8)$ R-symmetry. The corresponding transformations can be obtained
from the above $\textrm{SU}(2)\times \textrm{SU}(2)$ ones by performing there  the reduction (\ref{Red1}a). Note that the opportunity to obtain the $\cN{=}8$
supersymmetric $\tU(1)\times \tU(1)$ model through such a reduction of the ABJM $\textrm{SU}(2)\times \textrm{SU}(2)$ model was previously noticed in \cite{Li}.

Finally, we would like to point out that it is still an open question whether any other gauge $\cN{=}3$
Chern-Simons-matter
model with $\cN{=}8$ supersymmetry can be explicitly constructed. The $\cN{=}3$ harmonic formalism seems to be most appropriate
for performing such an analysis, since within its framework the issue of existence of one or another hidden symmetry amounts
to examining simple conditions under which unwanted quartic contributions to the full variation of the hypermultiplet parts
of the total action are vanishing.

\section{Discussion}
In this paper we gave a new superfield formulation of the ABJM theory with
gauge groups $\tU(N)\times \tU(N)$ and $\textrm{SU}(N)\times\textrm{SU}(N)$
as well as of some its generalizations,
in the harmonic $\cN{=}3, d{=}3$ superspace where three $d{=}3$ supersymmetries
are manifest and off-shell.
We found the $\cN{=}3$ superfield realization of the hidden $\cN{=}6$
supersymmetry and $\textrm{SO}(6)$ R-symmetry of the ABJM theory and
demonstrated how these symmetries are enhanced to $\cN{=}8$ and
$\textrm{SO}(8)$ in the BLG case of the gauge group
$\textrm{SU}(2)\times\textrm{SU}(2)$.
We also presented an example where $\cN{=}6$ supersymmetry and $\textrm{SO}(6)$
R-symmetry are reduced to $\cN{=}5$ and $\textrm{SO}(5)$, respectively.
One of the salient features of the $\cN{=}3$ formulation is that its
superfield equations of motion are written solely in terms of analytic
$\cN{=}3$ superfields and have a surprisingly simple form, see
\p{EqMo}, \p{Qeq} and~\p{VEqs}. Another nice property is that the invariant
actions are always represented by the difference of the $\cN{=}3$
superfield Chern-Simons actions for the left and right gauge groups plus
the actions of two hypermultiplets which sit in the bifundamental
representation of the gauge group and are minimally coupled to the gauge
superfields. No explicit superfield potential is present in the action,
as is dictated by the $\cN{=}3$ superconformal invariance.
The famous sextic scalar potential of the component formulation naturally
emerges on shell as a result of the elimination of some auxiliary
degrees of freedom from the gauge and hypermultiplet superfields.
The $\cN{=}3$ superfield formulation suggests a simple
technical criterion as to whether a chosen gauge group admits the existence
of hidden additional supersymmetries and R-symmetries: it is the cancellation
of the terms quartic in the hypermultiplets
in the full variation of the gauge-covariantized hypermultiplet action.

To clarify the significance of the $\cN{=}3$ superfield formulation
presented here, let us resort to the analogy between the ABJM theory and
the $\cN{=}4$, $d{=}4$ super Yang-Mills (SYM$^4_4$) theory, which describe
the low-energy dynamics of multiple M2 and D3 branes, respectively.
As is well known, the SYM$^4_4$ model is the maximally supersymmetric and
superconformal gauge theory in four dimensions, a fact crucial for
the string theory~/~field theory correspondence (see e.g.~\cite{AdsCFT}).
The $\cN{=}2$, $d{=}4$ harmonic superspace~\cite{GIKOS} provides the
appropriate off-shell $\cN{=}2$ superfield description of SYM$^4_4$ as
SYM$^2_4$ plus an $\cN{=}2$ hypermultiplet in the adjoint representation
minimally coupled to the $\cN{=}2$ gauge superfield.
Such a formulation was successfully used
to study the low-energy quantum effective action and
the correlation functions of composite operators in
$\cN{=}2$ superspace (see, e.g., \cite{d4} and~\cite{corr}).

Analogously to SYM$^4_4$, the ABJM model is the maximally supersymmetric
and superconformal Chern-Simons-matter theory in three dimensions \footnote{
Well, almost: The maximal supersymmetry in three dimensions with a highest spin
of one is $\cN{=}8$, corresponding to the BLG special case of the ABJM model.
However, the BLG model describes only {\it two\/} M2 branes since it is based
on $\textrm{SU}(2)\times\textrm{SU}(2)$ while the ABJM model serves perfectly
for an {\it arbitrary\/} number of M2 branes since it is based on
$\tU(N)\times \tU(N)$.}.
The ABJM construction opened up ways for studying the AdS$_4$/CFT$_3$
correspondence between three-dimensional field models and four-dimensional
supergravity in AdS space \cite{ABJM}--\cite{Ooguri}. We believe that the
$\cN{=}3$ superfield description of the ABJM model and its generalizations
developed in the present paper will be as useful for studying their algebraic
and quantum structure as the $\cN{=}2$ harmonic superspace approach has proved
to be for SYM$^4_4$. In particular, we expect that it will be
very efficient for investigating the low-energy quantum effective action in
three-dimensional $\cN{=}6$ supersymmetric field models as well as
for computing the correlation functions of composite operators directly
in $\cN{=}3, d{=}3$ harmonic superspace, because the manifest off-shell
$\cN{=}3$ supersymmetry is respected at each step of the computation.

Furthermore, there are also other directions
for extending the present study. A natural generalization
is to find the $\cN{=}3$ superfield description for
superconformal field models with $\cN{=}4$ and $\cN{=}5$
supersymmetries, which are also interesting from the point of view
of the AdS/CFT correspondence.
We already considered one such example in Subsection~4.3.
Another evident task is the coupling of the ABJM $\cN{=}3$ superfield models
to (conformal) $\cN{=}3$ superfield supergravity.

As one more possible development, one may hope that our $\cN{=}3$ superfield
re\-formulation is capable to give further insight into the structure of those BLG theories
which are based on the Nambu bracket (see \cite{NB}--\cite{Townsend}) and
to clarify their relation to the M5~brane. In this connection, we mention that
the equations of motion in the analytic $\cN{=}3$ superspace \p{Qeq} and
\p{VEqs} for the $\tU(N)\times \tU(M)$ model
(and their analogs for the other models considered) possess an equivalent
formulation in ordinary $\cN{=}3$ superspace as follows. Using the bridges
for the gauge superfields and passing to the central basis in $\cN{=}3$
harmonic superspace and the so-called $\tau$ gauge frame~\cite{book}, one can
convert the equations~\p{Qeq} to the form of flat harmonicity conditions,
which imply that the corresponding hypermultiplet superfields are linear in
the harmonics $u^{+}_i$. At the same time, the spinorial harmonic analyticity
conditions become highly nonlinear in this case, and one may think that the
$\tau$-frame form of the Chern-Simons equation~\p{VEqs} arises as an
integrability condition for these nonlinear harmonic analyticity constraints.
At this point there might be contact with a recent paper~\cite{Bandos},
where the equations of motion for the Nambu-bracket BLG theory were rewritten
in terms of $\cN{=}8$ superfields as some superfield constraint
of first order in a gauge-covariantized spinor derivative.
Based on the analogy with the ordinary $\cN{=}3$ superfield form of the ABJM
equations just mentioned, we guess that the constraint of~\cite{Bandos}
can be interpreted as a kind of Grassmann harmonic $\cN{=}8$ analyticity in
the $\tau$~frame.

Finally, it is worthwhile to note that the interrelations between the
low-energy actions describing M2 and D2 branes was the subject of many
papers (see, e.g.,~\cite{M2D2}). It was discovered that this issue is
intimately related to a new type of Higgs phenomenon. It is clearly of interest
to elaborate on it using our $\cN{=}3$ superfield framework.
In the Appendix we show how this phenomenon arises in the simplest
$\tU(1)\times \tU(1)$ model of Subsection~3.2.

\acknowledgments

I.B.S.~is very grateful to I.A.~Bandos, C.-S.~Park and D.P.~Sorokin,
for useful discussions and to INFN, Sezione di Padova \&
Dipartimento di Fisica ``Galileo Galilei'', Universita degli Studi
di Padova as well as to Institut f\" ur Theoretische Physik, Leibniz Universit\"at
Hannover for kind hospitality and support.
E.A.I.~is indebted to A.~Smilga
for interest in the work and useful discussions and to SUBATECH (Nantes)
and ENS (Lyon) for kind hospitality at the final stages of this study.
The present work is partly supported by an INTAS
grant, project No 05-1000008-7928, by RFBR grant, project
No 08-02-90490 and by a DFG grant, project No 436 RUS/113/669.
The work of I.L.B, N.G.P and I.B.S is supported also by RFBR
grant, project No 09-02-00078-a, and
by a grant for LRSS, project No 2553.2008.2.
The work of E.A.I. and B.M.Z. is partly supported by RFBR grant,
project No 06-02-16684-a, and by a grant
of the Heisenberg-Landau program.
I.B.S.\ acknowledges the support from an INTAS grant, project No
06-1000016-6108 and from the Alexander von Humboldt Foundation.
N.G.P.\ acknowledges the support from RFBR grant, project No 08-02-00334-a.

\appendix
\section{Appendix. Higgs effect in the $\tU(1)\times \tU(1)$ model}
\setcounter{equation}{0}
\renewcommand{\theequation}{A.\arabic{equation}}

Here we briefly discuss how the Higgs-type effect of refs.\ \cite{M2D2} arises in the
framework of the $\cN{=}3$ superfield formalism. We shall consider the simplest $\tU(1)\times \tU(1)$ model of Subsect.\ 3.2. The corresponding
superfield action, the sum of \p{VA} and \p{hyp22}, can be treated as a low-energy limit of the worldvolume action of single M2 brane.

The gauge group \p{u1u1HypG} which acts on hypermultiplets is realized by the following infinitesimal transformations
\be
\delta q^{+a} = \Lambda\, q^{+a}\,, \quad \delta \bar{q}{}^{+}_a  = -\Lambda\,\bar{q}{}^{+}_a\,, \quad
\delta A^{++} = - D^{++}\Lambda\,,\quad \Lambda = \Lambda_L - \Lambda_R\,.
\ee
The rest of the gauge $\tU(1)\times \tU(1)$ group with the parameter $\hat{\Lambda} =  \Lambda_L + \Lambda_R$,
acts only on $V^{++}$: $\delta V^{++} = -D^{++}\hat{\Lambda}\,$.

As the first step we pass in \p{hyp22} to the dual $\omega, f^{++}$ description by decomposing
\be
q^{+a} = u^{+a}\omega - u^{-a} f^{++}\,, \quad \bar{q}^{+}_a = -u^{+}_{a}\tilde{\omega} + u^{-}_a \tilde{f}^{++}\,.
\ee
Assuming that there is a constant real condensate in $\omega$,
\be
\omega = c_0 + \hat{\omega}\,, \quad \bar{c_0} = c_0\,,
\ee
and taking into account the gauge transformation law
\be
\delta \hat\omega = \Lambda \left(c_0 + \hat{\omega}\right),
\ee
one can choose the ``unitary'' gauge in which the imaginary part of $\hat\omega$ has been completely gauged away:
\be
\widetilde{\omega} = \omega\,, \quad \widetilde{\hat\omega} = \hat\omega\,.\lb{Ugau}
\ee
Up to a total harmonic derivative, the Lagrangian in the action \p{hyp22} in this gauge is rewritten as
\be
{\cal L}_q = \left(f^{++} + \tilde{f}^{++}\right) D^{++}\hat{\omega} -f^{++} \tilde{f}^{++}
-A^{++}\left(f^{++} - \tilde{f}^{++}\right)(c_0 +\hat{\omega}). \lb{H2}
\ee
Upon varying with respect to the auxiliary superfields $f^{++}, \tilde{f}^{++}$ and substituting the result
back into \p{H2}, we obtain
\be
{\cal L}_q \;\Rightarrow \tilde{\cal L}_q\; = (D^{++}\hat{\omega})^2 - (c_0+\hat{\omega})^2(A^{++})^2\,.
\ee
We see that the superfield $A^{++}$ is now also auxiliary and can be eliminated from the sum $S_{gauge} + S_{hyp}$, eqs.\ \p{VA}, \p{hyp22},
by using its algebraic equation of motion
\be
A^{++} = -\frac{ik}{16\pi}\,\frac{1}{(c_0+\hat{\omega})^2}\,W^{++}(V).
\ee
Substituting this expression back into the total action, we obtain
\be
S_{gauge} + S_{hyp} \;\Rightarrow \;
\int d\zeta^{(-4)} \left[(D^{++}\hat\omega)^2 - \frac{k^2}{(16\pi)^2} \frac{1}{(c_0+\hat{\omega})^2}W^{++}(V)W^{++}(V) \right].\lb{FinAct}
\ee
This action is a sum of the free real $\hat{\omega}$ hypermultiplet action and the ${\cal N}{=}3, d{=}3$ Maxwell action
multiplied by the ``dilaton'' factor which ensures the (spontaneously broken) superconformal invariance of the final
gauge-fixed action (recall that we started from the action invariant under the ${\cal N}{=}3, d{=}3$ superconformal symmetry).
It should also be implicitly invariant under nonlinearly realized $\textrm{SO}(6)$ symmetry and hidden ${\cal N}{=}3$ supersymmetry,
since these invariances  are  inherent in the sum of the actions \p{VA}, \p{hyp22} we started with. It is interesting
to inquire what kind of nonlinear sigma model for scalar fields is hidden in \p{FinAct}. One has now four real scalar fields in
the $\hat{\omega} $ hypermultiplet and three physical scalars in the gauge action (former auxiliary fields
$\phi^{(kl)}_L + \phi^{(kl)}_R$ of the Chern-Simons superfield action), i.e.\ total of seven physical scalar fields \footnote{
One missing scalar degree of freedom out of the initial eight ones is now described by the 3-dimensional
Abelian gauge field which is dual to a scalar field.}. One of these bosonic fields is dilaton, so there remain six bosonic fields
which should support a nonlinear realization of the group $\textrm{SO}(6) \sim \textrm{SU}(4)$. The only 6-dimensional coset manifold of $\textrm{SU}(4)$ seems
to be $\mathbb{CP}^3 \sim \textrm{SU}(4)/\tU(3)$, so we expect that the action \p{FinAct} contains the $d{=}3$ nonlinear $\mathbb{CP}^3$ sigma model in its bosonic
sector and thus can be interpreted as a low-energy limit of the single D2 brane action on $AdS_4\times \mathbb{CP}^3$.

It would be interesting to see how the above procedure generalizes to the non-Abelian  case.

\end{document}